\documentclass[11pt]{article}

\usepackage{./mystyle}
\begin{document}
\title{Configuration Space BPHZ Renormalization on Analytic Spacetimes}
\author{Steffen Pottel \thanks{Electronic address: steffen.pottel'at'yahoo.de}}
\affil{\small K\"uhne Logistics University, 20457 Hamburg, Germany }
\date{\today}
\maketitle
\begin{abstract}
A configuration space version of BPHZ renormalization is proved in the realm of perturbative algebraic quantum field theory. All arguments are formulated entirely in configuration space so that the range of application is extended to analytic spacetimes. Further the relation to the momentum space method is established. In the course of that, it is necessary to study the limit of constant coupling. 
\end{abstract}
\section{Introduction}

In the perturbative approach to quantum field theory on Minkowski space, almost all physical quantities are ill-defined already at finite order of the formal expansion. The pioneering work of Feynman, Tomonaga and Schwinger regarding loop corrections in quantum electrodynamics was studied more constructively by Dyson \cite{Dyson:1949bp} and led to a broad and intense development in the field of renormalization theory \cite{Velo:1976gh}, which vests perturbative quantum field theory with high predictive power. In the perturbative construction of interacting quantum field theories on curved spacetime, most of the prescriptions have to be reformulated since they rely heavily on techniques in momentum space. Instead, the Epstein-Glaser renormalization scheme \cite{Epstein:1973gw} was initially already constructed in configuration space and later extended to curved spacetimes \cite{Brunetti:1999jn,Hollands:2001nf,Hollands:2001fb} in the realm of algebraic quantum field theory. In that extension, the set of axioms, assessing whether a prescription is a renormalization scheme, was adapted and modified \cite{Hollands:2004yh} to be adequate for non-trivial geometries. Furthermore, the axioms admit a classification of the renormalization ambiguities, thus the conditions imposed on the equivalence of other prescriptions to the Epstein-Glaser scheme. Subsequently, methods have been developed for Mellin-Barnes regularization \cite{Hollands:2010pr} (requiring specific spacetimes), dimensional regularization on flat configuration space \cite{Bollini:1972ui,'tHooft:1972fi,Duetsch:2013xca} and analytic regularization on curved spacetimes \cite{Speer:1972wz,Gere:2015qsa}. It is worth noting that all of them resolve the combinatorial structure with forest formulas. \\
The forest formula was introduced to renormalization theory in the BPHZ renormalization method \cite{Bogoliubov:1957gp,Hepp:1966eg,Zimmermann:1968mu,Zimmermann:1969jj} resolving the combinatorial problem arising from renormalization parts, which cannot be regularized simultaneously. The regularization itself, called $R$-operation, is a variation of the Hadamard regularization in the sense that, instead of defining it on test functions in dual spaces, the Taylor subtraction is computed directly on a weighted Feynman graph. The BPHZ scheme and its modification BPHZL \cite{Lowenstein:1975rg,Lowenstein:1975ps,Lowenstein:1975ku}, required if additionally massless fields are present in a theory, were used, for instance, in the BRST quantization \cite{Becchi:1975nq,Tyutin:1975qk}, the definition of composite operators \cite{Zimmermann:1972te} or for the proof of Wilson's operator product expansion in perturbation theory \cite{Wilson:1972ee,Zimmermann:1972tv}.\\
It is the objective of the present work to combine the structural advantages of the algebraic approach in perturbative quantum field theory and of the BPHZ prescription in the realm of renormalization. In particular, we want to enlarge the applicability of BPHZ renormalization to curved spacetimes, for which it is natural to formulate the entire scheme in configuration space, and the following heuristic argument supports the possibility of such a formulation. Restricted to Minkowski space and the vacuum state, the (inverse) Fourier transform of the propagator or fundamental solution can be computed explicitly and, after the application of the $R$-operation, the weight of a Feynman graph is a well-defined tempered distribution in momentum space, thus its inverse Fourier transform exists formally. With a proper definition of the $R$-operation, i.e.\ the Taylor operator of the Hadamard regularization, it is reasonable to derive a renormalization prescription in configuration space, which follows the common thread of BPHZ renormalization but is independent of the momentum space prescription.\\
The paper is organized as follows. After introducing all necessary notions for the formulation regarding the extension problem in renormalization in the next section, we derive a renormalization prescription and prove equivalence to the Epstein-Glaser method in Section 3. In the fourth and fifth section we provide a sufficient condition for the existence of the constant coupling limit and compare the additional subtractions stemming from differing definitions of renormalization parts, respectively. Finally, we discuss our results and set them into perspective of future research.

\section{Preliminaries}

The main result of this work is derived for a four-dimensional globally hyperbolic analytic spacetime $(M,g)$, where $g$ is a Lorentzian metric with signature $(+,-,-,-)$. In particular, the Minkowski space $(\IR^4,\et)$, used in the comparison of our result to the momentum space methods, is an example of such a spacetime.\\
As motivation for the upcoming construction, consider a single real scalar field $\vph$ and a potential term $\CL_I(\vph) = -V(\vph)= \CO(\vph^3)$, which fulfills the equation of motion
\begin{align}\label{eq:ELnl}
	P \vph = -\frac{\de V(\vph)}{\de \vph}, 
\end{align}
where $P$ is a normal hyperbolic differential operator of second order. Since there is no general well-posedness theory available for \eqref{eq:ELnl}, we apply a standard argument from perturbation theory and study the field $\vph$ expanded about the exactly solvable free field $\ph$ fulling the linear equation of motion
\begin{align}\label{eq:freeKG}
	P\ph(x)=0 .
\end{align}
For \eqref{eq:freeKG}, it is possible to find local advanced/retarded fundamental solutions $F_\pm^\Om$ in a geodesically convex region $\Om$, using the Hadamard parametrix construction \cite[Chapter 2]{Bar:2007zz}, where only the spacetime geometry and the parameters in $P$ are included in. The Hadamard parametrix $H_\pm$ takes the form
\begin{align}
	H_\pm(x,y) = \frac{1}{4\pi^2} \left[ \frac{U(x,y)}{\si_\pm(x,y)} + V(x,y) \log\left(\frac{\si_\pm(x,y)}{\La}\right) \right],
\end{align}
where $V$ is a formal power series with finite radius of convergence and the index $\pm$ at the squared geodesic distance $\si$ denotes
\begin{align}
	\si_\pm(x,y) \bydef \si(x,y) \pm i\epsilon (T(x)-T(y)) + \frac{\epsilon^2}{4},
\end{align}
with $T$ being the global time function, so that $H_\pm$ is a well-defined distribution in the limit $\epsilon\rightarrow 0$. Since we work with analytic spacetimes, $H$ coincides with the exact fundamental solution $F_\pm$ restricted to $\Om$ \cite{Friedlander:2010eqa} and due to the global hyperbolicity of the spacetime, local fundamental solutions can be glued together resulting in a global fundamental solution such that \eqref{eq:freeKG} has smooth global solutions $\ph\in C^\infty(M)$ for the Cauchy problem \cite[Chapter 3]{Bar:2007zz}
\begin{align}
	\left\lvert 
	\begin{array}{ll}
	P u = f               & \mbox{ on }M     \\
	u_{|\Si} = u_0 \\
	\na_\n u_{|\Si} = u_1 & \mbox{ on } \Si, 
	\end{array}
	\right. 
\end{align}
where $\Si$ is a Cauchy surface, $f\in\CD(M)$ and $u_0,u_1\in\CD(\Si)$. Anticipating the quantum character, we promote the field to a distribution, informally expressed by 
\begin{align}
	\ph(f) = \int_M\limits \ph(x) f(x) d\m, \quad f\in\CD(M),
\end{align}
and use it to generate the free, unital $*$-algebra $\SCA(M,g)$, which satisfies the conditions
\begin{align}
	&\ph(f)^* - \ph(\ol f) = 0, \nonumber \\
	\label{eq:WeakFieldEquation}
	&\ph(Pf) = 0, \\
	&\ph(a f_1 + b f_2) - a \ph(f_1) - b \ph(f_2) = 0 \mbox{ with } a,b\in \IC, \nonumber \\
	&[\ph(f_1),\ph(f_2)] - iF(f_1,f_2) \II = 0, \nonumber
\end{align}
where $F$ is the commutator function defined as the difference of advanced and retarded fundamental solution. Elements $\ph(f)\in\SCA(M,g)$ are considered to be random variables which give meaningful information only after taking the expectation value, i.e acting with a state
\begin{align}
	\om: \SCA(M,g) \rightarrow \IC 
\end{align}
on the fields. $\om$ is a linear map, which	satisfies the normalization $\om(\II)=1$ and the positivity condition $\om(\ph(f)^* \ph(f))\geq 0$ for all $\ph(f)\in\SCA$. A state is said to be Hadamard \cite{Kay:1988mu} if for a geodesically convex region $\Om$ the two-point function is given by
\begin{align}
	\om(\ph(x),\ph(y)) = H(x,y) + W_{\om}(x,y),
\end{align}
where $x,y\in\Om$ and $W_\om$ is smooth, such that the singularity structure is completely determined by the Hadamard parametrix $H$, thus only by the geometry and the parameters in $P$. With this property, we can define Wick ordering independently of the state, which is referred to as locally covariant constructed\cite{Brunetti:2001dx}. All elements of $\SCA(M,g)$ may be expressed recursively by
\begin{align}\label{eq:WickOrdering}
	:\ph(f):_H                           & \bydef \ph(f)\nonumber\\
	:\ph(f_1)...\ph(f_n):_H \ph(f_{n+1}) & = :\ph(f_1)...\ph(f_{n+1}):_H\\
	                                     & \qquad + \sum_{j=1}^n :\ph(f_1)...\widecheck{\ph(f_j)}...\ph(f_n):_H H(f_j,f_{n+1}),\nonumber 
\end{align}
where $\widecheck\bullet$ denoted the extraction of that field from the Wick polynomial. Therefore the product of two Wick polynomials is again expressed recursively by Wick polynomials \cite{Hollands:2001nf}, i.e.\ since Wick ordering was defined symmetrically, we have in a geodesically convex region $\Om\subset M$
\begin{multline}
	:\ph(x_1)...\ph(x_n):_H \cdot :\ph(y_1)...\ph(y_m):_H =\\
	= \sum_{k\leq\min(n,m)} (-1)^k \sum_{\{(i,j)\}_k} :\ph(x_1)...\widecheck{\ph(x_{i_1})}...\widecheck{\ph(x_{i_k})}...\ph(x_n) \times \\
	\times \ph(y_1) ... \widecheck{\ph(y_{j_1})} ... \widecheck{\ph(y_{j_k})} ... \ph(y_m):_H \prod_{l=1}^k H(x_{i_l},y_{j_l}),
\end{multline} 
where $\{(i,j)\}_k$ denotes $k$ mutually disjoint pairs $(i_l,j_l)$ with $l=1,...,k$, $i_l=1,...,n$ and $j_l=1,...,m$. We further remark that the Wick ordering can be carried out independently of covariant derivatives acting on the field $\ph$, factors constructed locally covariant from the metric and constants of the theory like mass $m$ or coupling to curvature $\x$. Let us denote by 
\begin{align}
	\CP(x) \bydef \CP[g_{ab}, R_{abcd}, \na_{(e_1}...\na_{e_k)}R_{abcd},\x,m^2](x)
\end{align}
a polynomial in the metric $g$, the Riemann tensor $R_{abcd}$, its symmetrized covariant derivatives as well as the mass $m$ and the coupling $\x$ to the curvature so that a generalized Wick monomial can be written as 
\begin{align}
	\Ph(x) \bydef \CP(x) \prod \na_{(f_1}...\na_{f_l)}\ph(x).
\end{align}
Indeed, it follows from the Thomas Replacement Theorem \cite{Hollands:2007zg} that locally covariant Wick monomials may only depend on elementary fields, its covariant derivatives as well as elements in $\CP$, which allows us to define the algebra of field observables
\begin{align}
  	\SCB(M,g) \bydef \left\{ :\Ph:_{H} (f) | f\in\CD(M) \right\}. 
\end{align}
From this point on, we omit writing the subscript $H$ and consider the normal ordering always in the sense of Hadamard. For the perturbative construction of interacting quantum field theories, we further require the notion of time-ordered products. At this stage, we additionally perform a transition to the off-shell formalism, i.e.\ factors $P\ph(f)$ in elements of $\SCB(M,g)$ do not fulfill the weak field equation \eqref{eq:WeakFieldEquation}. Naive time-ordering $\CT$ of elements in the algebra of field observables $\SCB(M,g)$ is defined via
\begin{align}
	\CT(:\Ph(x):) & \bydef :\Ph(x): \nonumber ,\\
	\label{eq:NaiveTimeOrdering}
	\CT(:\Ph(x): \cdot :\Ph(y):) & \bydef 
	\begin{cases} 
     	:\Ph_1: (x) \cdot :\Ph_2: (y) \quad \mbox{  for  } \quad x\notin J^-(y)\\
     	:\Ph_2: (y) \cdot :\Ph_1: (x) \quad \mbox{  for  } \quad y\notin J^-(x),
   	\end{cases}
\end{align}
where no particular order is preferred if $x$ and $y$ are acausally separated, and all higher orders are defined recursively. We want to relate the time ordering to the Wick product of \eqref{eq:WickOrdering}. For simplicity, we consider the product of two fields in a geodesically convex domain $\Om\subset M$.
\begin{align}
	\CT(\ph(x)\ph(y)) & = :\ph(x)\ph(y): + 
	\begin{cases} 
     	H_-(x,y) \quad \mbox{  for  } \quad x\notin J^-(y)\\
     	H_+(x,y) \quad \mbox{  for  } \quad y\notin J^-(x),
   	\end{cases}
\end{align}
Using the global time function $T$, we define the Feynman Hadamard parametrix
\begin{align}\label{eq:FeynmanHadamard}
	H_F(x,y) \bydef \te(T(x)-T(y)) H_+(x,y) + \te(T(y)-T(x)) H_-(x,y),
\end{align}
where $\te(\bullet)$ denotes the Heaviside step-function. Since the product $uv$ of two distributions $u,v\in\CD'(M)$ can be defined as the pullback of the tensor product $u\otimes v$ by the diagonal map $\de$ if for their analytic wavefront sets
\begin{align}
	(x,k)\in\WF_\mathrm{A}(u) \quad\Rightarrow\quad (x,-k)\notin \WF_\mathrm{A}(v)
\end{align}
holds for some $(x,k)$ \cite[Section 8.5]{hormander1990analysis}, we read off from \eqref{eq:FeynmanHadamard} that $H_F\in\CD'(\Om\times \Om \setminus \mbox{"diagonal"})$. Introducing the notion of UV-scaling degree of a distribution $u\in\CD'(\IR^n\setminus\{0\})$ as
\begin{align}
    \uvs(u) \bydef \inf\left\{\al\in\IR| \lim_{\la\rightarrow0}\limits \la^\al u_\la = 0 \right\} 
\end{align}
where 
\begin{align}
  \langle u_\la,f\rangle & \bydef \langle u,f^\la\rangle \\
  f^\la     & \bydef \la^{-n} f(\la^{-1}\bullet),
\end{align}
the distribution $u$ or the Feynman propagator can be uniquely extended if $\uvs(u) < n$ or $\uvs(H_F) < \dim(M)$, respectively \cite[Thm. 5.2]{Brunetti:1999jn}. Looking at a local version of Wick's theorem \eqref{eq:WickOrdering}
\begin{align}
	T & (:\Ph_1(x_1): ... :\Ph_n(x_n):) = :\Ph_1(x_1) ... \Ph_n(x_n): + \nonumber\\
	& + \sum_{(i,j);i<j} \na_{ij}H_F(x_i,x_j) :\Ph_1(x_1) ... \Ph^{(1)}_i(x_i) ... \Ph^{(1)}_j(x_j) ... \Ph_n(x_n): \nonumber \\
	& + (\mbox{higher orders}) \nonumber\\
	= & \sum_{\al_1,...,\al_n} \frac{1}{\al_1!...\al_n!} \bigotimes_{(i,j);i<j} (\na_{ij}H_F)^{a_{ij}}(x_i,x_j) :\Ph_1^{(\al_1)}(x_1) ... \Ph_n^{(\al_n)}(x_n):, \label{eq:LocalWickHF}
\end{align}
with $\Ph^{(\al)}$ denoting the $\al$-th functional derivative and $\na_{ij}$ covariant derivatives stemming from the definition of $\Ph$, we observe that the extension problem becomes significantly more involved, when arbitrary Wick monomials are considered since we have to find a prescription such that the tensor product can be defined as a pointwise product and, only with this, we can define a mechanism which extends $\prod (\na H_F)^a$ to a distribution over $\CD(M^n)$. A priori, it is not clear that any constructed extension is physically reasonable, i.e.\ it is a coherent prescription for all time-ordered products of field monomials $\Ph(f)$. In \cite{Hollands:2001nf,Hollands:2001fb} and later extended in \cite{Hollands:2004yh}, the authors give a set of axioms, which assess whether a chosen regularization and extension prescription is physically reasonable and show that such a prescription exists. Without going into the details, the main criteria regard causality, unitarity and covariance. Further appropriate scaling behavior under rescalings of the metric and the microlocal spectrum condition, a generalization of the Hadamard condition for two-point functions, are demanded. Only those prescriptions fulfilling the axioms are referred to as \emph{renormalization schemes}. The construction of renormalization schemes is not unique, but it can be shown that different prescriptions are equivalent. The idea goes back to Hepp \cite{Hepp:1969bn} and was picked up in \cite{Hollands:2001nf,Hollands:2002ux,Hollands:2007zg}, which states that two schemes are equivalent if their time ordered products can be related by a finite change in the ambiguities of the extension described above. Specifically, let us denote the ambiguities by $\De\in\CE'(M^n)$ with $\supp(\De)\subset \{(x_1,\ldots,x_n)\in M^n| \exists i,j\in\IN, 1\leq i< j \leq n, \mbox{ such that } x_i = x_j \}$. Then we require that $\De$ is constructed locally covariant and scales almost homogeneously, i.e.\ $\De$ scales homogeneously up to logarithmic corrections. In contrast to the UV-scaling degree, the scaling of $\De$ is determined by the engineering dimension and thus includes curvature terms and parameters in the wave operator $P$. It follows that the ambiguities depend polynomially on the field $\ph$, the mass parameter $m^2$ and Riemann curvature tensor. Furthermore $\De$ should be symmetric in its arguments and real. If the ambiguities $\De$ of time-ordered product $\CT$ have the properties described above, then $\CT$ defines a new renormalization scheme satisfying the axioms provided $\CT$ can be related to another renormalization scheme $\hat \CT$ by \cite[Thm. 2]{Hollands:2007zg}
\begin{multline}\label{eq:PresEquiv}
	\hat \CT\{ :\Ph_1(x_1): ... :\Ph_n(x_n):\}  = \CT\{:\Ph_1(x_1): ... :\Ph_n(x_n):\} \\
	+ \sum_{c\geq 1} \sum_{\{\SV\}_c} \CT \left\{ \bigotimes_{k\in\SV_0} :\Ph_k(x_k): \bigotimes_{l=1}^c \De_{\SV_l} \left( \bigotimes_{l'\in\SV_l} :\Ph_{l'}(x_{l'}): \right) \right\},
\end{multline}
where $\SV_0 \cup \{\SV\}_c = \{1,...,n\}$ and $\SV_i\cap\SV_j = \emptyset$. One may rephrase this statement in the following way. Two definitions of the time-ordered products $\CT$ and $\tilde \CT$ are equivalent renormalization schemes if $\tilde\CT$ is a renormalization scheme and they can be related by a finite redefinition, a \emph{renormalization}, of $\CT$. \\
In the original formulation of BPHZ renormalization \cite{Bogoliubov:1957gp,Hepp:1966eg,Zimmermann:1968mu,Zimmermann:1969jj}, the renormalization of a single naively defined time-ordered product of Wick monomials is given in momentum space by applying Bogoliubov's $R$-operation to numerical distributions, which are derived by Wick's theorem. The combinatorial structure behind those numerical distributions and the recursive action of the $R$-operation may be better understood in terms of so-called Feynman graphs and is resolved by the forest formula. In the following, we transfer the approach to a prescription elaborated entirely in configuration space. In particular, it allows for a transition to non-trivial analytic spacetimes. The construction of the BPHZ scheme in configuration space is performed in three steps. First, we introduce a special prescription of analytic continuation of the metric so that the $R$-operation can be carried out on the numerical distribution kernel. Second, we prove that the forest formula solves the underlying combinatorial problem of overlapping divergences. Finally, we remove the analytic continuation and show that our construction indeed defines a renormalization scheme. We remark that we do not distinguish among various choices of equivalent Wick monomials \cite[Thm. 5.1]{Hollands:2001nf}. For one thing, we are mostly interested in the dynamical objects, the elementary fields $\ph$, since those are responsible for the restriction of the domain, and for another thing the choice differs by linear combinations of Wick monomials, thus simply leading to further independent extension problems.

\section{Convergence of the $R$-Operation}

For the configuration space formulation of BPHZ renormalization, we begin with a naively (in the sense of \eqref{eq:NaiveTimeOrdering}) defined time-ordered product
\begin{align}\label{eq:timeorderedMain}
	\CT\left\{:\Ph_1(f_1): \cdot ... \cdot :\Ph_n(f_n):\right\}
\end{align}
with $:\Ph_j(f_j):\in\SCB(M,g)$ and $\supp(f_i)\cap\supp(f_j) = \emptyset$. After the application of Wick's theorem \eqref{eq:LocalWickHF} restricted to a geodesically convex region $\Om\subset M$, we obtain the numerical distribution
\begin{align}
	v_o \bydef \bigotimes_{(i,j); i<j} (\na_{ij}H_F)^{a_{ij}}(x_i,x_j)\in\CD'(((\Om\times \Om)\setminus \diag)^{\sum a_{ij}}).
\end{align}
We note that each $H_F$ may be interpreted as a graph with two vertices and one edge. Let us subsume those in an abstract edge set $E$ such that $v_0$ is expressed by the $|E|$-fold tensor product over Feynman parametrices $H_F$. We further observe that the time-ordered product \eqref{eq:timeorderedMain} depends on $n$ arguments $x_i$ before smearing with test functions $f_i$. In particular, these arguments are the only available arguments in the factors $H_F$ of $v_o$ so that we subsume them in an abstract vertex set $V$. This identification gives rise to the following definition.
\begin{definition}\label{de:feyng}
  A \emph{Feynman graph} $\G(E,V)$ with $n$ vertices of valency one, called external, and $k$ vertices of valency strictly larger than one, called internal, consists of two finite sets $V$ and $E$ together with a map $\pa : E \rightarrow V\times V / \sim$, where $\sim$ is the equivalence relation $(a, b)\sim(b, a)$, called incidence map such that if $e\in E$, $\pa(e)=\{a, b\}$ with $a,b\in V$. \\
  If $\G(E,V)$ is a directed graph, then $\pa e=(s(e),t(e))$ is an ordered pair with $s,t:E\rightarrow V$.
\end{definition}
With this, we indicate the graph structure of the numerical distribution by $v_0\bydef v_0[\G]$ and find that
\begin{align}
	\CT\left\{:\Ph_1(x_1): \cdot ... \cdot :\Ph_n(x_n):\right\} = \sum_\G v_0[\G] :\Ph(x_1)...\Ph(x_n)[\G]:,
\end{align}
where $:\Ph(x_1)...\Ph(x_n)[\G]:$ denotes the resulting Wick product after applying the necessary contractions, i.e.\ functional derivatives, as in \eqref{eq:LocalWickHF}. Since the total number of graphs for a single time-ordered product of Wick monomials is finite, it suffices to restrict our considerations to a single Feynman graph $\G$. 

In the next step, we relate the distribution $v_0[\G]$ over the edge set $E(\G)$ to a distribution $u_0[\G]$ over the vertex set. Recalling from the Definition \ref{de:feyng} that the boundary operator $\pa$ maps elements in $E$ to elements in $V$, we would like to establish 
\begin{align}\label{eq:PullbackDist}
	u_0[\G] = d^\ast v_0[\G],
\end{align}
using the coboundary operator
\begin{align}
	d:V(\G) \rightarrow E(\G).
\end{align}
Since each $H_F$ is a distribution, \eqref{eq:PullbackDist} is not naively defined. However, it becomes well-defined if we can find a regularization $v_0^\e[\G]$ such that its wavefront sets admit the pointwise product. In particular, the pointwise product becomes well-defined if the regularization $\e$ is chosen such that the projection to the first variable of the wavefront set $\WF_\mathrm{A}(H_F)$, i.e.\ the singular support of $H_F$, is contained in the diagonal. Suppose we find such a regularization, then 
\begin{align}
	u_0^\e[\G] \bydef d^\ast v_0^\e[\G] \in \CD'(\Om^{|V(\G)|}\setminus\mbox{"graph contractions"}).
\end{align}
Those ``graph contractions'' describe configurations in which connected subgraphs are contracted to a point, i.e.\ at least one edge $e\in E(\G)$ lies on the thin diagonal $\diag$ and we want to define the set of "graph contractions" as the graph diagonal.
\begin{definition}
  Let $\G(V,E)$ be a Feynman graph. Then the \emph{large graph diagonal} is defined by
  \begin{align}
    \oo & \bydef \{x\in M^{|V|} | \exists \g\subset\G \mbox{ connected } \forall v,w\in V(\g), v\neq w: x_v=x_w \} \\
    \intertext{and the \emph{thin graph diagonal} by}
    \OO & \bydef \{ x \in M^{|V|} | \forall v,w\in V(\G): x_v=x_w \}.      
  \end{align}
\end{definition}
Next, we turn to the construction of a regularization of the Hadamard parametrix in the spirit of Zimmermann \cite{Zimmermann:1968mu}. Since $(M,g)$ is globally hyperbolic, we exploit that, due to the isomorphism $\Ps: M \rightarrow \IR \times \Si$, the metric $g$ can be written as $\be dt^2 - g_t$, where $g_t$ is Riemannian. Additionally we assumed $(M,g)$ to be analytic so that there exists a unique analytic continuation 
\begin{align}\label{eq:MetricAnaCont}
	g^\e \bydef (1-i\e) \be dt^2 - g_t,\quad \e>0,
\end{align}
of the metric. This continuation is sufficient to render the pointwise product well-defined. Specifically, we prove in the first step that there exist Riemannian bounds on $g^\e$.  
\begin{lemma}\label{le:MetricEstimate}
	Let $g^\e$ be given by \eqref{eq:MetricAnaCont} and define
	\begin{align}
		g^R \bydef \be dt^2 + g_t.
	\end{align}
	For every $x\in\Om\subset M$, $\Om$ geodesically convex, and every $\x\in T_x\Om$
	\begin{align}
		\hat C(\e) g_R(\x,\x) \leq |g^\e(\x,\x)| \leq \check C(\e) g_R(\x,\x)
	\end{align}
	holds, where
	\begin{align}
		\hat C(\e) & = \left(\frac{1}{\e} + \sqrt{1+\frac{1}{\e^2}}\right)^{-1},\\
		\check C(\e) & = \sqrt{1+\e^2}.
	\end{align}
\end{lemma}
\begin{proof}
	Consider any $x\in \Om$ and any $\x\in T_x\Om$. We compute
	\begin{align}
		\frac{|g^\e_x(\x,\x)|^2}{(g^R_x(\x,\x))^2} = \frac{(\be \x_0^2 - g_t(\x,\x))^2}{(\be \x_0^2 + g_t(\x,\x))^2} + \frac{\e^2 \x_0^4 \be^2}{(\be \x_0^2 + g_t(\x,\x))^2} \leq 1+\e^2.
	\end{align}
	This proves the second inequality of the assertion. For the first inequality, we write
	\begin{align}
		\frac{g^R_x(\x,\x)}{|g^\e_x(\x,\x)|} & = \frac{\be \x_0^2}{|\be \x_0^2 - g_t(\x,\x) - i\e \be \x_0^2|} \nonumber\\
		& \hspace{1cm} + \frac{g_t(\x,\x)}{\sqrt{\be^2 \x_0^4 - 2\be \x_0^2 g_t(\x,\x) + (g_t(\x,\x))^2 + \e^2 \be^2 \x_0^4}} \nonumber\\
		& \leq \frac{1}{\e} + \frac{g_t(\x,\x)}{\sqrt{(1+\e^2-\al) \be ^2 \x_0^4 + \left( 1 - \frac{1}{\al} \right) (g_t(\x,\x))^2 }} \label{eq:MetricYoung}\\
		& = \frac{1}{\e} + \frac{g_t(\x,\x)}{\sqrt{\left( 1- \frac{1}{1+\e^2} \right) (g_t(\x,\x))^2}}
		= \frac{1}{\e} + \sqrt{1+\frac{1}{\e^2}}, \nonumber
	\end{align}
	where we used Young inequality to get \eqref{eq:MetricYoung} and set $\al=1+\e^2$ afterwards.
\end{proof}
This result is the analogue of the Euclidean estimates in \cite{Zimmermann:1968mu}. While Lemma \ref{le:MetricEstimate} is sufficient for the Fourier transform of propagators in Minkowski space, we require another argument such that $d^\ast v_0[\G]$ becomes well-defined. Recall that the Hadamard parametrix $H$ was constructed as a local fundamental solution to the differential operator $P$. After the analytic continuation of the metric, also the differential operator changes accordingly, since we assumed $P$ to be normal. Let us denote this by $P_\e$. Further we observe that the Hadamard parametrix depends purely on geometric data, thus is constructed with respect to $g^\e$ so that
\begin{align}
	P_\e H_\e = \de.
\end{align} 
The properties of $H_\e$ are sufficient to render the pointwise product well-defined.
\begin{proposition}
	Let $g^\e$ be an analytic continuation of the metric given by \eqref{eq:MetricAnaCont} and $\Om\subset M$ be geodesically convex. Then $u_0^\e[\G] \bydef d^\ast v_0^\e[\G]$ is well-defined and
	\begin{align}
		u_0^\e[\G] \in \CD'(\Om^{|V(\g)|}\setminus\oo).
	\end{align}
\end{proposition}
\begin{proof}
	Since $H_\e$ is a local fundamental solution to $P_\e$, we obtain by microelliptic regularity \cite[Thm. 8.6.1]{hormander1990analysis} that
	\begin{align}
		\WF_\mathrm{A}(H_\e) \subseteq \Char (P_\e) \cup \WF_\mathrm{A}(\de),
	\end{align}
	where $\de$ denotes the Dirac-$\de$-distribution and $\Char(P_\e)$ is the characteristic set of $P_\e$, i.e.\ the points $(x,k)\in T^\ast\Om$ for which the principal symbol $\si_{P_\e}$ vanishes excluding the zero section. With the estimates of Lemma \ref{le:MetricEstimate}, we note that the characteristic set of $P_\e$ is empty since $g^R$ is Riemannian. Thus 
	\begin{align}
		\WF_\mathrm{A}(H_\e) \subseteq \WF_\mathrm{A}(\de).
	\end{align}
	Moreover we have $H_\e\in\CD'(\Om\times\Om\setminus\diag)$ so that products of regularized Hadamard parametrices and therefore $d^\ast v_0^\e[\G]$ are well-defined. Furthermore, we note that $u_0^\e[\G]$ is not defined if edges, thus connected subgraphs, are contracted to a point, which coincides with the definition of the large graph diagonal $\oo$.
\end{proof}
We turn to the problem of extending the graph weight $u_0^\e[\G]\in\CD'(\Om^{|V(\G)|}\setminus\oo)$. The idea of Bogoliubov and Parasiuk was to introduce an $R$-operation, i.e.\ one replaces the distribution $u_0^\e[\g]$ with $\g\subseteq \G$ by its Taylor remainder in order to meet the requirement on the UV-scaling degree for the unique extension \cite[Thm. 5.2]{Brunetti:1999jn}. In the following, we call a graph $\g$ divergent or renormalization part if its weight does not fulfill the necessary constraint on the UV-scaling degree. Defining this $R$-operation recursively throughout the full graph $\G$ by assigning a subtraction degree to each subgraph determining the order of Taylor subtraction, one ends up with a distribution extended to the whole space in the ideal case, i.e.\ in the case of non-overlapping divergent graphs.
\begin{definition}
  Two graphs $\g$ and $\g'$ are \emph{overlapping}, denoted by $\g \olap \g'$, if none of the following conditions 
  \begin{align}\label{eq:overlapping}
    V(\g)\subseteq V(\g'), \quad V(\g)\supseteq V(\g'), \quad V(\g)\cap V(\g') = \emptyset 
  \end{align}
  hold. Otherwise they are non-overlapping, denoted by $\g\nolap\g'$.
\end{definition}
We note that Zimmermann \cite{Zimmermann:1969jj} defines overlap with respect to the edge set $E$. This mismatch to our definition results from the change of relevant variables when transferring from momentum space to configuration space. In the momentum space treatment, one associates the momenta to flows through lines rather than to vertices, which account only for momentum conservation. Instead, the relative position of vertices adjacent to the same edge determines their correlation in configuration space. Therefore it is sufficient to restrict the set of renormalization parts to full vertex parts, i.e.\ graphs $\g$ with $V(\g)$ and all edges connecting these vertices. 
The problem of such overlapping graphs may be resolved by collecting all divergent parts in a family of partially ordered sets. For our purpose, those are sets of subgraphs $\g\subseteq\G$ together with the usual inclusion $\subseteq$. Zimmermann introduced in \cite{Zimmermann:1969jj} the notion of forests, which are made up of all sets of non-overlapping graphs $\g\subseteq\G$.
\begin{definition}
  A $\G$-forest $F$ is a partially ordered set (poset) over $V(\G)$, where for each pair of elements in $F$ one relation of \eqref{eq:overlapping} holds.
\end{definition}
Note that the condition on subgraphs is less restrictive than in \cite{Zimmermann:1969jj}, hence gives rise to possibly more renormalization parts. Nevertheless we proceed to follow the idea of Zimmermann, i.e.\ in contrast to the initial $R$-operation, one does not apply the full Taylor operation to the distribution, i.e.\ computing always the Taylor remainder, but assigns to each element $f$ of a $\G$-forest $F$ the corresponding Taylor polynomial.
\begin{definition}\label{de:taylor}
  	Let $f\in C^k(\Om)$ for $\Om\subset\IR^{n}$ convex. For $d\leq k$ and multiindex $\al$ with $|\al|\leq d$, the Taylor polynomial of $f$ about a point $\ol x$ is given by
  	\begin{align}
  		t^{d}_{x}|_{\ol{x}} f(x) \bydef \sum_{|\al|=0}^{d}\limits \frac{(x-\ol{x})^\al}{\al!} f^{(\al)}(\ol x).
  	\end{align}
\end{definition}
We choose the point of subtraction to be located at the thin graph diagonal of the renormalization part. For any graph $\g\subseteq\G$, its thin graph diagonal depends on the configuration of $\g$ in space, i.e.\ on $x_v\in\IR^d$ for $v\in V(\g)$, and by this the point of subtraction is not a constant but variable. We set $\ol{V(\g)}$ to be the vertex which is computed by
\begin{align}\label{eq:CMCoordinate}
	x_{\ol{V(\g)}} \bydef \frac{1}{2|E(\g)|} \sum_{v\in V(\g)} |E(\g|v)| x_{v},
\end{align}
where $E(\g|v)$ denotes the set of incident edges at vertex $v\in V(\g)$ contributing to $\g$. We remark that Steinmann \cite[Section 10.3]{Steinmann:2000nr} defines the point of subtraction to be the standard mean coordinate
\begin{align}
	\ol x = \frac{1}{|V(\g)|} \sum_{v\in V(\g)} x_{v}.
\end{align}
While both points of subtraction may be used for the definition of BPHZ renormalization in configurations space, it turns out that \eqref{eq:CMCoordinate} is necessary for the derivation of normal products in the sense of Zimmermann \cite{Pottel:2017cc}. 

Note that for an edge weight $u_0^\e[e]=H_\e$ with $e\in E(\G)$, the mean coordinate coincides with the thin diagonal. Hence the Taylor operator cannot be applied directly to that weight. More generally, consider a graph $\G$ such that $\g\subset \G$ and denote the mean coordinate of $\g$ by $\ol{V(\g)}$. We write formally $u_0^\e[\G] = u_0^\e[\G\lineco\g] u_0^e[\g]$, where $\lineco$ denotes the line complement of $\g$ such that the sets of arguments of both factors are not disjoint. While $u_0^\e[\g]$ becomes singular at $\ol{V(\g)}$, we demand $u_0^\e[\G\lineco\g]$ to be smooth in a neighborhood of $\ol{V(\g)}$. Furthermore it carries all arguments connecting $\G\lineco\g$ to $\g$. Hence the application of the Taylor operator $t^{d(\g)}_{V(\g)}|_{\ol{V(\g)}}$ to $u_0^\e[\G\lineco\g]$ is defined and it remains to show that this prescription yields the desired properties as suggested by the original BPHZ scheme. 
\begin{definition}
  Let $\G$ and $\g\subset\G$ be graphs with weights $u_0^\e[\G]$ and $u_0^\e[\g]$, respectively. Then we define the action of the operator $\IP(\g)$ by
  \begin{align}
    t^d_{x|\ol x} \IP(\g) u_0^\e[\G] \bydef u_0^\e[\g]\, t^d_{x|\ol x} u_0^\e[\G\lineco\g] \,.
  \end{align}
  Here, the set difference $\lineco$ is meant to be computed with respect to the set of lines. In the case of $\IP(\G)$, $\IP$ maps only to the vertex weights $\prod u_0^\e[v]$. If $u_0^\e[\G]$ does not contain any vertex weights, we employ standard Hadamard regularization on test functions in the dual space of $u_0^\e[\G]$.
\end{definition}
\begin{remark}
    The operator $\IP$ is not a projection operator. It reorders the distributional kernel in such a way that the action of the Taylor operation is well-defined and thus may be viewed as the counterpart of Zimmermann's substitution operator $S_\g$. Recall that $S_\g$ assigned momenta in $\g$ such that the Taylor polynomial is always computed at zero external momenta of $\g$. In the same sense, $\IP(\g)$ ensures that the Taylor polynomial can be computed at the thin graph diagonal of $\g$.
\end{remark}
In order to determine the necessary degree of the Taylor polynomial, we have to look at two competing mechanisms. On the one hand there is the scaling degree, which quantifies how fast the weight diverges near the graph diagonal. On the other hand, the scaling can be viewed as a continuous change in the configuration, i.e.\ the embedding of the graph into the spacetime. Evidently one can reach the graph diagonal by keeping one vertex fixed and contracting edge after edge to a point. Then the continuous change in configurations turns into integrations, since graph weights are functionals, and we end up with the notion of the UV-degree of divergence 
\begin{align}
	\uvd(u_0^\e[\g]) \bydef \uvs(u_0^\e[\g]) - \dim(M)(|V(\g)|-1).
\end{align}
for a weight $u_0^\e[\g]\in\CD'(\Om^{|V(\g)|}\setminus\oo)$. With this we collected all necessary ingredients for the definition of the configuration space forest formula in the sense of Zimmermann.
\begin{definition}\label{de:ForestFormula}
	Let $\G(V,E)$ be a Feynman graph and $u_0^\e[\G]\in\CD'(\Om^{|V(\G)|}\setminus\oo)$ be the smooth weight over $\G$. The \emph{$R$-operation} on the graph weight is given by
  	\begin{align}\label{eq:forestformula}
    	Ru_0^\e[\G] \bydef \sum_{F\in\SF}\limits \prod_{\g\in F} (- \underbrace{t^{d(\g)}_{V(\g)}|_{\ol{V(\g)}} \IP(\g)}_{t(\g)\textrm{ for short}}) u_0^\e[\G], 
  	\end{align}
  	where $\SF$ is the set of all $\G$-forests, $d(\g) \bydef \lfloor \uvd(u_0^\e[\g]) \rfloor$ and the Taylor operators are ordered in the sense that $t(\g)$ appears left of $t(\g')$ if $\g\supset \g'$ and no order is preferred if $\g\cap\g'=\emptyset$.
\end{definition}
Note that the point, about which the Taylor expansion is performed, is not fixed in spacetime but moves according to changes of the configuration of the graph, thus remains variable. 
With the analytic continuation of the metric and the forest formula, we defined all components of the BPHZ renormalization scheme in configuration space.
\begin{theorem}\label{th:Convergence}
	Let $\G(V,E)$ be a Feynman graph and $u_0^\e[\G]\in\CD'(\Om^{|V(\G)|}\setminus\oo)$ be the weight over $\G$. Then $Ru_0^\e[\G]$ can be uniquely extended to $Ru^\e\in \CD'(\Om^{|V(\G)|})$ and 
  	\begin{align}
  		\lim_{\e\rightarrow0}\langle Ru^\e[\G], f\rangle = \langle Ru^\e[\G] , f\rangle
  	\end{align}
  	converges for all $f\in\CD(\Om^{|V(\G)|})$. $R$-operation and naive time-ordering $\CT$ define a renormalization scheme.
\end{theorem}
\begin{remark}
	Since there are no further assumptions on the parameters in the wave operator $P$, the construction of the Hadamard parametrix $H_\e$ holds for any mass parameter $m$. Thus the result of Theorem \ref{th:Convergence} holds for both massive and massless scalar quantum fields.
\end{remark}
Since we established the relation to the Riemannian metric in Lemma \ref{le:MetricEstimate}, the equivalent statement follows in the Riemannian case.
\begin{corollary}\label{co:Riemann}
	Consider $(\Om,g^R)$ and let $\G(V,E)$ be a Feynman graph and $u_0[\G]\in\CD'(\Om^{|V(\G)|}\setminus\oo)$ be the smooth weight over $\G$. Then $Ru_0[\G]$ can be uniquely extended to $Ru\in \CD'(\Om^{|V(\G)|})$.
\end{corollary}
For the proof of Theorem \ref{th:Convergence}, we recall that a distribution is uniquely extendible if the UV-scaling degree $\uvs(\bullet)$ is smaller than the space dimension. Since the weights $u_0^\e[\g]$ are analytic in the neighborhood of any graph diagonal, the condition on the UV-scaling degree can be equivalently rephrased in the sense that the weights can be uniquely extended if they are locally integrable in a neighborhood of the graph diagonal. Therefore extendability follows from local integrability in a region $\Om'^{|V(\G)|}\subset\IR^{4|V(\G)|}$, where $\Om'$ is mapped diffeomorphically to $\Om\subset M$ via the exponential map. 
\begin{theorem}[Thm. 1, \cite{Pottel:2017aa}]\label{th:L1loc}
	Let $\CK[\G]\in\ C^\infty(\IR^{d|V(\G)|}\setminus\oo)$ be the weight over a simple graph $\G$, which has positive scaling degree at the large graph diagonal. Then 
	\begin{align}
		R\CK[\G] \in L^1_\mathrm{loc}(\IR^{d|V(\G)|}) \, .
	\end{align}
\end{theorem}
Let us give the idea of the proof of Theorem \ref{th:L1loc}, which follows the original argument in \cite{Zimmermann:1969jj} closely and requires essentially two steps.
First, the forests are reordered with respect to a vertex set $I \subset V(\G)$ such that the operator $(1-t(\g))$, with $\g \subseteq \G$, is applied to $\CK[\G]$ if $|V(\g) \setminus I| = 1$ and $- t(\g)$ otherwise.
We observe that the condition $|V(\g) \setminus I| = 1$ corresponds to integrating out all but one vertex of $\g$ and, thus, the Taylor remainder is computed exactly for those renormalization parts, which are integrated over the thin graph diagonal.
In particular, since the reordering is defined for general sets $I\subset V(\G)$ it follows that $R\CK[\G]$ can be adapted subsequently to $I \cup \{v\}$ for any additional element $v\in V(\G) \setminus I$ and therefore to any sequence of integrations.
Second, it is verified that the introduced Taylor operators entail the desired regularizing effect.
Indeed, for $\la, \g \subset \G$ it follows that
\begin{align}
	\uvs_{I_\g}( - t(\la) \IP(\la) \CK[\G] ) 
	& \leq \uvs_{I_\g} ( \CK[\g] ) \\
	\uvs_{I_\g}( (1 - t(\g)) \IP(\g) \CK[\G] ) 
	& \leq \uvs_{I_\g} ( \CK[\g] ) - d(\g) - 1 \, ,
\end{align}
where positive scaling degree at the large graph diagonal is necessary if $\la \olap \g$.
These relations are then utilized in a recursive argument.
In order to illustrate the latter, consider a set $I \subset V(\G)$, a renormalization part $\g \subset \G$, and a reordered $\g$-forest $F$.
Further let $\g_1,\ldots,\g_a \in F$ be maximal subgraphs of $\g$, i.e.\ there exists no element $\g'\in F$ with $\g_i \subset \g' \subset \g$.
Denoting the restriction of the $R$-operation to $\g$ and $\g_1,\ldots,\g_a$ by $R_{F,\g}$, the inequality
\begin{align}
	\uvs_{I_\g} (R_{F,\g} \CK[\G]) < d |I_\g|
\end{align}
holds, provided the same relation holds for $\g_1,\ldots,\g_a$ separately as well.
With this, the recursion starts at minimal renormalization parts for each $\G$-forest $F$, where minimal renormalization parts do not contain subgraphs which are renormalization parts themselves.
Proceeding with the recursion until $\G$ is reached, the bound
\begin{align}
	\uvs_I (R \CK[\G] ) < d|I|
\end{align}
is obtained, which implies local integrability of $R\CK[\G]$.

Furthermore, we notice the transition from a Feynman graph, i.e.\ a multigraph, to a simple graph in the statement of Theorem \ref{th:L1loc}. Therefore we need to justify that reducing the complexity of a graph by allowing maximally one edge to connect a given pair of vertices is still sufficient. This holds because we assign a positive but arbitrary scaling degree to each edge. Hence multiples of parametrices $H_\e$ turn out to be more rigid than the assumptions in Theorem \ref{th:L1loc} in fact. 
Further, the number of edges among two vertices is basically blind towards the calculation of Taylor polynomials and, in account with \eqref{eq:CMCoordinate}, we may replace the weight in the definition of the subtraction point by
\begin{align}
	x_{\ol{V(\g)}} = \frac{1}{\uvs(\CK[\g])} \sum_{v\in V(\g)} \frac{\uvs(\CK[e|v])}{2} x_v,
\end{align}
where $\CK[e|v]$ denotes the edge weights incident to $v\in V(\g)$. The difference between multigraphs and simple graphs may amount to combinatorial factors, but in particular not to a different behavior in the scaling. We pick up this transition in the discussion of additional subtractions appearing in the momentum space approach to BPHZ renormalization.

With the result of Theorem \ref{th:L1loc}, we are in the situation of having negative degree of divergence near any thin graph diagonal. The extension to the whole space then amounts to applying the extension to the thin diagonal of all (sub-)graphs iteratively.
\begin{lemma}\label{le:extenstion}
  	The $R$-modified weight $Ru_0^\e[\G]\in\CD'(\Om^{|V(\G)|}\setminus\oo)$ can be uniquely extended to $Ru^\e[\G]\in\CD'(\Om^{|V(\g)|})$.
\end{lemma}
\begin{proof}
  	We know from Theorem \ref{th:L1loc} that
  	\begin{align}\label{eq:ExtensionProof}
  		\uvs_I(Ru_0^\e[\G|_\g]) = \uvs_{I_\g}(Ru_0^\e[\G|_\g]) < \dim(M) |I_\g|.
  	\end{align}
  	Thus near each thin graph diagonal, the weight can be uniquely extended. Choose any $v_1\in V(\G)$ and determine the maximal variable subgraphs $\ta_{1i}$. The corresponding $Ru_0^\e[\G|_{\ta_{1i}}]$ is defined up to the thin diagonal of $\ta_{1i}$, but has a unique extension due to \eqref{eq:ExtensionProof}. Assume that all maximal variable subgraphs for $\{v_1,...,v_N\} \bydef I_N\subset V(\G)$ are uniquely extended. Then determine all maximal variable subgraphs $\ta_{N+1,i}$ with respect to $I_{N+1}$, where $I_{N+1}\subset V(\G)$ and $I_{N+1}\setminus I_N = \{v_{N+1}\}$. Note that $\ta_{N+1,i}$ is independent of the sequence $I_1\subset...\subset I_N$, i.e.\ they are the same for any permutation $\{v_{\p(1)},...,v_{\p(N)}\}$. Hence $\ta_{N+1,i}$ are defined up to the thin diagonal and we use \eqref{eq:ExtensionProof} again for the unique extension. Iterating until $I_\mathrm{max}\subset V(\G)$ with $I_\mathrm{max}=|V(\G)|-1$ proves the assertion.
\end{proof}
The last step to show convergence of the configuration space BPHZ method is the removal of the $\e$-regularization of the Hadamard parametrix $H_\e$ or equally the $\e$-dependence of the metric $g^\e$.
\begin{lemma}
	Let $Ru^\e[\G]\in \CD'(\Om^{|V(\G)|})$ be the weight over a graph $\G$. Then 
	\begin{align}
		\lim_{\e\rightarrow 0} Ru^\e[\G] = Ru[\G] \in \CD'(\Om^{|V(\G)|}).
	\end{align}
\end{lemma}
\begin{proof}
	For the proof of this Lemma, we want to use Theorem 3.1.15 \cite{hormander1990analysis}, which states that, in the limit of vanishing $\e$-regularization, $Ru^\e[\G]$ converges to a distribution in $\CD'(\Om)$ (including the extension to the whole region) provided we can bound $Ru^\e[\G]$ by some inverse monomial of the imaginary part of its arguments, where the imaginary part has to be an element of an open convex cone. In a geodesically convex $\Om\subset M$, we can express $H_F$ in its local form and find
	\begin{multline}
		\prod_{(i,j); i<j} H_\e^{a_{ij}}(x_i,x_j) \simeq \prod_{(i,j); i<j} \left( \frac{U_\e(x_i,x_j)}{\si_\e(x_i,x_j)} \right)^{a_{ij}} \\+ \mbox{ logarithmic corrections of lower order}.
	\end{multline}
	We observe that $\sum_{(i,j);i<j} a_{ij} = |E(\G)|$ so that, for our purpose, it is convenient to work w.l.o.g. with one edge of multiplicity $|E(\G)|$, i.e.\
	\begin{align}
		u_0^\e[\G] \simeq \frac{\Ps_\e(x,y)}{(\si_\e(x,y))^{|E(\G)|}} + \mbox{ log. corr. },
	\end{align}
	where $x,y\in(\Om\setminus\oo)$ and $\Ps$ is smooth. In the next step, we include the derivatives coming from the $R$-operation. Neglecting the moments of the Taylor polynomials, we get
	\begin{align}
		Ru^\e[\G] \simeq \sum_{F\in\SF} \left( \frac{\Ps_\e(x,y)}{(\si_\e(x,y))^{|E(\G)|}} \right)^{\left(\sum_{\g\in F} d(\g) \right)} + \mbox{ log. corr. }
	\end{align}
	Denoting the maximum of derivatives by
	\begin{align}
		d_\mathrm{max} \bydef \max_{F\in\SF} \sum_{\g\in F} d(\g)
	\end{align}
	and taking only those derivatives into account, which act on the denominator, the leading order is proportional to
	\begin{align}
		(\si_\e(x,y))^{-(|E(\G)|+d_\mathrm{max})}. 
	\end{align}
	We observe that, due to our choice of analytic continuation, we have
	\begin{align}
		\Im(\si_\e(x,y)) = -\e (T(x)-T(y))^2,
	\end{align}
	where $T(\bullet)$ is the time function in $(M,g^\e)$. For the application of Theorem 3.1.15 \cite{hormander1990analysis}, we have to consider arguments of each edge $e\in E(\G)$, which we subsume in one vector $\Si\in\IC^{|E(\G)|}$, where we informally write
	\begin{align}
		\Si = \left(\Re(\si_{\e,e_1}) + i\Im(\si_{\e,e_1}),...,\Re(\si_{\e,e_{|E(\G)|}}) + i\Im(\si_{\e,e_{|E(\G)|}})\right)^T.
	\end{align}
	such that
	\begin{align}
		|\Im(\Si)| = \e \sqrt{\sum_{e\in E(\G)} (\De T(e))^4},
	\end{align}
	where $\De T(e)$ denote the difference of the global times at the vertices of edge $e$.
	Further we find that $\Im(g^\e_x(\x,\x))\in V^-$, which is an open convex cone. Thus we conclude by
	\begin{align}
		|Ru^\e[\G]| & \simeq |\Ps_\e \cdot (\si_\e^{-(|E(\G)|+d_\mathrm{max})})| + \mbox{lower order terms} \\
		& \lesssim (\hat C(\e) \si^R)^{-(|E(\G)|+d_\mathrm{max})} \\
		& \lesssim \left( \frac{1}{\e} + \sqrt{1+\frac{1}{\e^2}} \right)^{(|E(\G)|+d_\mathrm{max})} \\
		& \lesssim \e^{-(|E(\G)|+d_\mathrm{max})} \\
		& \lesssim |\Im(\Si)|^{-(|E(\G)|+d_\mathrm{max})}.
	\end{align}
	Note that for the case of $\Si=0$, we obtain $Ru^\e[\G] = Ru[\G]$. Hence we are able to apply Theorem 3.1.15 \cite{hormander1990analysis} and find that $Ru^\e[\G]$ converges in $\CD'(\Om)$ in the limit of vanishing $\e$. 
\end{proof}
As a last step, we have to identify the ambiguities. Consider any renormalization part $\g\subseteq\G$. In the neighborhood of any graph diagonal, we may decompose the forest formula after saturation into contributions of forests with overlap and forests enforcing the Taylor remainder
\begin{align}
  R_\G u_0^\e[\G] = \Big( \underbrace{\sum_{F_{\mathrm{ol}}\in\SF_{\mathrm{ol}}} \prod_{\g'\in F_\mathrm{ol}} (-t(\g'))}_{\bydef X_\mathrm{ol}} + \underbrace{\sum_{F_\mathrm{non}\in\SF_\mathrm{non}} \prod_{\g\in F_\mathrm{non}} (-t(\g))}_{\bydef X_\mathrm{non}} \Big) u_0^\e[\G].
\end{align}
We know that $X_\mathrm{ol}u_0^\e[\G]$ is smooth at $\ol \g$ and thus focus on $X_\mathrm{non}u_0^\e[\G]$. We compute
\begin{align}
  X_\mathrm{non}u_0^\e[\G] & = \sum_{\ol F_\g\in \ol \SF_\g} \prod_{\si\in\ol F_\g} (-t(\si)) (1-t(\g)) \sum_{\ul F_\g\in\ul \SF_\g} \prod_{\si'\in\ul F_\g} (-t(\si')) u_0^\e[\G] \\ 
  & = \sum_{\ol F_\g\in \ol \SF_\g} \prod_{\si\in\ol F_\g} (-t(\si)) (1-t(\g)) u_0^\e[\G\lineco\g] R_{\g'} u_0^\e[\g] \\
  & \simeq \sum_{|\al|=d(\g)+1} \frac{1}{\al!} R_{\G/\g} (D^\al_{\ol\g} u_0^\e[\G/\g]) (x-\ol x_\g)^\al R_{\g'} u_0^\e[\g], 
\end{align}
where the sum over all forests containing $\g$ can be split into all $\g$-subforest $\ul F_\g$, with $\si'\subset \g$ for $\si'\in\ul F_\g$, and all $\g$-superforests $\ol F_\g$, with $\si\supset \g$ or $\si\cap\g=\emptyset$ for $\si\in \ol F_\g$. We observe that $u_0^\e[\g]$ can be changed by terms which are supported only in $\ol x_\g$ and scale maximally with order $d(\g)$. It is well-known \cite[Chapter 2]{hormander1990analysis} that the only distributions supported in a point are Dirac-$\de$-distributions and derivatives thereof. Therefore $u_0^\e[\g]$ is determined only up to the addition of
\begin{align}
  u_0^\e[\ol \g] \bydef \sum_{|\al|=0}^{d(\g)} c_\al(\ol x_\g) D^\al \de_{\ol x_\g},
\end{align}
where the coefficient function $c_\al(\ol x_\g)$ is fixed after employing suitable normalization conditions. Recall from \eqref{eq:PresEquiv} that those ambiguities may additionally relate our construction to another renormalization scheme. For this purpose, we introduce the $W$-projection from Epstein-Glaser renormalization \cite{Epstein:1973gw,Brunetti:1999jn}, which utilizes usual Hadamard regularization in the sense that, including the dual pairing, we write
\begin{align}
	\langle u_0^\e[\G], W f\rangle,
\end{align}
where $f\in\CD(\Om^{|V(\G)|})$. It is important to note that new test functions $w$ get introduced in the definition of the Taylor operator, i.e.\
\begin{align}
	Wf \bydef (1-t[w](\g)) f(x) \bydef f(x) - \sum_{|\be|=0}^{d(\g)} w_\be D^\be f(\ol x_\g)
\end{align}
for a single renormalization part $\g$.  We remark that the Epstein-Glaser method may be employed in the treatment in Feynman graphs, but is usually formulated more generally, and observe the similarity in the type of subtractions. Consequently, we compare
\begin{align}
	\langle (1-t(\g))u_0^\e[\G] + u_0^\e[\G/\g] u_0^\e[\ol\g]), f\rangle
\end{align}
to
\begin{align}
	\langle u_0^\e[\G], Wf\rangle
\end{align}
and compute
\begin{align}
	\langle (1-t(\g))u_0^\e[\G], f\rangle & = \langle (1-t(\g))u_0^\e[\G], Wf + t[w](\g)f\rangle \nonumber\\
	& = \underbrace{\langle u_0^\e[\G],Wf \rangle}_{\mbox{EG}} - \underbrace{\langle t(\g)u_0^\e[\G],Wf\rangle}_{\mbox{EG finite}} \nonumber\\
	& \hspace{2cm} + \underbrace{\langle (1-t(\g)) u_0^\e[\G], t[w](\g) f\rangle}_{\mbox{BPHZ finite}}. \label{eq:EquivTermsEgBphz}
\end{align}
Hence our method can be transferred into the Epstein-Glaser method if we renormalize the ambiguities so that they equal the second and third term in \eqref{eq:EquivTermsEgBphz}, where the former is finite using the Epstein-Glaser prescription and the latter is finite using our BPHZ prescription. Spelling out each term, i.e.\
\begin{align}
	\langle u_0^\e[\G/\g] u_0^\e[\ol\g], f\rangle & = \left\langle u_0^\e[\G/\g] \sum_{|\al|=0}^{d(\g)} c_\al(\ol x_\g) D^\al \de_{\ol x_\g}, f \right\rangle, \\
	\langle t(\g)u_0^\e[\G],Wf\rangle & = \left\langle \sum_{|\be_1|=0}^{d(\g)} \frac{(x-\ol x_\g)^{\be_1}}{\be_1!} D^{\be_1}_{\ol V_\g} u_0^\e[\G/\g] u_0^\e[\g], Wf \right\rangle, \\
	\langle (1-t(\g)) u_0^\e[\G], t[w](\g) f\rangle & = \left\langle (1-t(\g)) u_0^\e[\G], \sum_{|\be_2|=0}^{d(\g)} w_{\be_2} D^{\be_2} f_{\ol \g} \right\rangle,
\end{align}
we obtain
\begin{align}\label{eq:EquivOnGraphLevel}
	\langle u_0^\e[\G], Wf\rangle = \langle (1-t(\g))u_0^\e[\G] + u_0^\e[\G/\g] u_0^\e[\ol\g]), f\rangle
\end{align}
setting
\begin{multline}
	\left\langle u_0^\e[\G/\g] c_\al(\ol x_\g) D^\al \de_{\ol x_\g}, f \right\rangle = \\
	= \left\langle \frac{(x-\ol x_\g)^{\al}}{\al!} D^{\al}_{\ol V_\g} u_0^\e[\G/\g] u_0^\e[\g], Wf \right\rangle - \left\langle (1-t(\g)) u_0^\e[\G], w_{\al} D^{\al} f_{\ol \g} \right\rangle .
\end{multline}
We note that \eqref{eq:EquivOnGraphLevel} may be viewed as an intermediate step of \eqref{eq:PresEquiv}, which concludes Theorem \ref{th:Convergence} if we can sum over all contributing graphs. 

Due to the chosen regularization by the $R$-operation, we are able to derive a condition on $u_0^\e[\ol \g]$. Recall that $\G/\g$ is the reduced graph in which $\g$ is contracted to a vertex $\ol V$. Let us denote by $\ol \SE_{\ol V}$ and $\ol\SD_{\ol V}$ the set of elementary field operators and the set of covariant derivatives, respectively, which are assigned to external lines of $\g$ and to the vertex set $V(\g)$, thus incident lines of $\ol V$ after the contraction of $\g$. With this, we obtain
\begin{align}\label{eq:AmbiguityInFields}
  u_0^\e[\ol \g] = \sum_{|\al|=0}^{d(\g)} c_\al(\ol x_\g) D^\al \prod_{k\in\ol\SE} \na_{(k)} \ph (\ol x_\g).
\end{align}
We remark that those counterterms $u_0^\e[\ol \g]$, which we may associate to each renormalization part in $\G$, inherit all locally covariant terms which remain from the the initially considered monomials in the time-ordered product, i.e.\ all vertex weights of $\g$. Recall that those terms are important for the engineering dimension but not for the UV-scaling degree. Furthermore, $u_0^\e[\ol \g]$ scales almost homogeneously, since it is an insertion with dimension $d(\ol \g) = |\ol\SE_{\ol V}| + |\ol\SD_{\ol V}| + |\al|$ into another time-ordered product restricted to the graph $\G/\g$. 

Next we sum over all graphs $\G$ analogously to \cite{Clark:1976ym}. For this purpose, consider the sets of maximal non-overlapping renormalization parts  over $\G$ in $\SF$. For each set $\{\g_1,...,\g_c\}$, the forests can be subsumed to $\g_j$-forests. To each $\g_j$, we can assign a vertex set $\SV_j$ and a set $\SE_j$ associated to elementary fields $\ph$ constructing the edges of $\g_j$. Obviously, $(\SV_j,\SE_j)$ does not determine $\g_j$ uniquely. Note that the complement $\ol\SE_j$ is uniquely defined for each pair $(\SV_j,\SE_j)$ and the number of elementary fields in each counterterm gets fixed. For some $\G$, $\G/\g_1...\g_c$ is the reduced graph with $c$ new vertices. We decompose
\begin{align}
  V(\G) = \bigcup_{j=1}^c \SV_j \cup \ol \SV.
\end{align}
Including the ambiguities informally on the level of graphs, we obtain
\begin{align}
  \G \mapsto \G + \sum_{c\geq1}\sum_{\{\g_j\}_c} \G/\g_1...\g_c,
\end{align}
where each $\G/\g_1...\g_c$ has $\ol V_1,..., \ol V_c$ new vertices. The dimension of each new vertex $\ol V_j$ is given by
\begin{align}
	d(\ol V_j) \bydef |\ol\SE_j| + |\ol\SD_j| + |\al_j|.
\end{align}
Note that the counterterm graphs are recursively related to each other, i.e.\ by the dimension constraint and the structure of the $R$-operation, all reduced graphs may be further reduced in subsequent applications of the $R$-operation and result in new ambiguities. But those only exist for supergraphs which are already renormalization parts. Hence the ambiguities are defined recursively in accordance to the definition of the $R$-operation. Denoting by $\{D\}_c$ the derivatives stemming from Taylor operators, we compute
\begin{align}
  \sum_\G Ru_0^\e[\G] & \mapsto \sum_\G Ru_0^\e[\G] + \sum_\G \sum_{c\geq 1} \sum_{\{\g\}_c} \sum_{\{D\}_c} Ru_0^\e[(\G/\g_1...\g_c)\cup \ol\g_1...\ol\g_c] \nonumber\\
  & = \sum_\G Ru_0^\e[\G] + \sum_{c\geq 1} \sum_{\{\SV\}_c} \sum_{\{\SE\}_c} \sum_{\{\g\}_c} \sum_{\{D\}_c} \sum_{\ol \G_c} Ru_0^\e[(\G/\{\g\}_c)\cup\{\ol\g\}_c]. \label{eq:EquivSumGammasIntermediate}
\end{align}
Going the inverse direction in Wick's theorem, we have
\begin{align}
  \sum_\G Ru_0^\e[\G] = \CT_{R,\e}\left\{ \prod_{j=1}^n \Ph_n(f_n) \right\}.
\end{align}
For the second term in \eqref{eq:EquivSumGammasIntermediate}, we observe that the sum over all sets $\{\g\}_c$ was required to determine the subtraction degree $d(\g_j)$ of each $\g_j$ via the UV-degree of divergence. In fact, we find
\begin{align}
	d(\g_j) = 2|E(\g_j)| - 4(|V(\g_j)|-1) + |\na|
\end{align}
for scalar fields with dimension one in four spacetime dimensions with $|\na|$ denoting the number of covariant derivatives acting on edges in $\g$ and neglecting possible improvements of the degree due to vertex weights. But, by definition, 
\begin{align}
	|\SE_j|=2|E(\g_j)|, \quad |\SV_j| = |V(\g_j)| \quad \mathrm{and} \quad |\SD_j|=|\na|
\end{align}
holds for any $\g_j$ so that
\begin{align}\label{eq:GeneralGraphDegree}
	d(\g_j) = 4 + |\SE_j| - 4 |\SV_j| + |\SD_j|
\end{align}
and the sum over all sets $\{\g\}_c$ can be performed independently of the sets $\{D\}_c$. We may absorb the result in the coefficients $c_{\al_j}(\ol x_{\SV_j})$ such that we obtain
\begin{multline}\label{eq:AmbiguitiesSummedUp}
	\sum_{c\geq 1} \sum_{\{\SV\}_c} \sum_{\{\SE\}_c} \sum_{\{\g\}_c} \sum_{\{D\}_c} \sum_{\ol \G_c} Ru_0^\e[(\G/\{\g\}_c)\cup\{\ol\g\}_c] = \\
	= \sum_{c\geq 1} \sum_{\{\SV\}_c} \sum_{\{\SE\}_c} T_{R,\e} \Bigg\{ \prod_{v\in\ol\SV} \Ph_v(f_v) \prod_{j=1}^c  \sum_{|\al_j|=0}^{d_j} c_{\al_j}(x_{\ol V_j}) D^{\al_j}_{\ol V_j} \ol \SE_j(f_{\ol V_j}) \Bigg\}
\end{multline}
with $d_j$ given by \eqref{eq:GeneralGraphDegree} and $\ol\SE_j$ understood, in the sense of \eqref{eq:AmbiguityInFields}, as monomial in $\ph$ and its covariant derivatives $\na$. We notice that we arrive at the desired relation, i.e.\ the ambiguities can be expressed by locally covariant field monomials inserted into time-ordered products. Hence they are supported on the diagonal and fulfill the scaling constraint. By the definition of the subtraction point $x_{\ol V_j}$, the counterterms are symmetric in their arguments, counting every elementary field operator $\ph$, and, by construction of $\ph\in\SCA(M,g)$, they are real. In particular, \eqref{eq:AmbiguitiesSummedUp} converges to a well-defined distribution in the limit $\e\rightarrow 0$. This concludes the proof of Theorem \ref{th:Convergence}.

\section{Limit of Constant Coupling}

In the remaining part of this work, we want to establish a relation of our results from the previous section to the momentum space method \cite{Zimmermann:1968mu,Zimmermann:1969jj}. For this purpose, we restrict our considerations to Minkowski space $(\IR^4,\et)$, where, without analyzing the problem in momentum space in great detail, we are confronted with the problem whether our construction still holds if the test functions assigned to inner vertices of a Feynman graph are replaced by a constant. Recall that we constructed a renormalization scheme for weighted Feynman graphs $u_0[\G]$ on analytic spacetimes $(M,g)$, where the edge weights $u_0[e]$, $e\in E(\G)$, were only specified by their UV-scaling degree. For the results of this part, we would like to further specify them with respect to their long-range behavior. For a general distribution $u\in\CD'(\IR^n)$, its scaling was defined in the weak sense
\begin{align}
	(u_\la,f) = (u,f^\la),
\end{align}
i.e.\ via the scaling of a test function $f\in\CD(\IR^n)$. For long ranges, we turn to the ''inverse`` case and consider again the scaling $u_\La$ of a distribution $u\in\CD'(\IR^n)$. At large values of $\La$, it is not reasonable to work in the weak sense due to the compact support of the test functions. Hence we require additionally that $u$ is a regular distribution, i.e.\ $u=T_f$ with $f\in L^1_{\mathrm{loc}}$. This admits the following definition.
\begin{definition}
	Let $u\in\CD'(\IR^n)$ be a regular distribution. Then the \emph{large argument scaling} of $u$ is defined by
	\begin{align}
		\irs(u) \bydef \sup \left\{ \al\in\IR| \lim_{\La\rightarrow \infty} \La^\al u_\La = 0 \right\}
	\end{align}
	with $u_\La = u(\La x)$. The IR-degree of divergence is given by
	\begin{align}
		\ird(u) \bydef \irs(u) - n.
	\end{align}
\end{definition}
Denoted in this way, UV- and IR-degree of divergence are notationally inverse to the definition of Lowenstein and Zimmermann \cite{Lowenstein:1975rg}, but, of course, do not change the notion. In fact, the change of notation is very natural, considering that large frequencies correspond to small wavelengths after Fourier transformation so that the underline in $\uvd$ and the overline in $\ird$ are associated to small and large values, respectively, regardless of configuration or momentum space.

In the conventional approach to quantum field theories \cite{Itzykson:1980rh}, one usually defines the interaction with respect to a coupling constant. Up to this stage, we considered all monomials (associated to vertices in a graph) as algebra-valued distributions, thus any ``coupling'' was represented by a compactly supported smooth function. In order to be able to relate our result from Theorem \ref{th:Convergence} to the results of Zimmermann \cite{Zimmermann:1968mu,Zimmermann:1969jj}, we have to let test functions for internal vertices of a connected graph approach a constant, i.e.\ we are concerned with the question whether
\begin{align}\label{eq:ConstCoupLimTProduct}
	\lim_{g_j\rightarrow \mathrm{const.}} T_{R,\e}^\mathrm{conn} \left\{ \prod_{i=1}^m \Ph^\mathrm{lin}_i(f_i) \prod_{j=1}^n \Ph_j^\mathrm{nlin}(g_j) \right\}
\end{align}
exists, where the $\Ph^\mathrm{lin}_i$ are only linear in the field $\ph$ and the $\Ph_j^\mathrm{nlin}$ are strictly nonlinear in the field $\ph$ such that the former correspond to external vertices $V_\mathrm{e}(\G)$ and the latter correspond to internal vertices $V_\mathrm{i}(\G)$ for a connected graph $\G$ with vertex set $V(\G)\bydef V_\mathrm{e}(\G) \sqcup V_\mathrm{i}(\G)$. The idea for the existence of the limit lies in standard real analysis. Namely, we may bound the evaluation of a distribution $u\in\CD'(\IR^n)$ together with any test function $f\in\CD(\IR^n)$ by
\begin{align}
	\langle u,f \rangle = \int_{\IR^n} u(x) f(x) dx \leq \|f\|_\infty \int_{\IR^4} u(x) dx \leq \|f\|_\infty  \|u\|_1
\end{align}
provided $u$ is integrable. With this, the limit $f\rightarrow \mathrm{const}$ exists such that we find a constraint on the IR-scaling degree of the distribution is sufficient for integrability.
\begin{theorem}\label{th:ConstCoupLim}
	Let $Ru_0^\e[\G]\in\CD'(\IR^{4|V(\G)|}\setminus\oo)$ be the weight over a Feynman graph $\G$. If 
	\begin{align}\label{eq:ConstCoupLimAssumption}
		\irs_I(u_0^\e[\G]) > 4|I|
	\end{align}
	for any $I\subseteq V_\mathrm{i}(\G)$, then the limit of constant coupling
	\begin{align}
		\lim_{g\rightarrow\mathrm{const}} \langle Ru_0^\e[\G], f\otimes g \rangle 
	\end{align}
	in the sense of \eqref{eq:ConstCoupLimTProduct} exists.
\end{theorem}
We remark that the assumption refers to the weight $u_0^\e[\G]$ without being modified by the $R$-operation. Before we prove that this is sufficient for the limit to exist, let us further motivate the condition. 
\begin{proposition}\label{pr:MinkIFF}
	The weight $Ru_0^\e[\G]$ on Minkowski spacetime $(\IR^4,\et)$ is absolutely integrable for any $I\subseteq V_\mathrm{i}(\G)$ if and only if the weight $Ru_0^E[\G]$ on Euclidean space $(\IR^4,\de)$ is absolutely integrable for any $I\subseteq V_\mathrm{i}(\G)$.  
\end{proposition}
Suppose for now that it is sufficient to control the unmodified kernel $u_0^\e[\G]$. The factors of $u_0^\e[\G]$ in Minkowski space are given by Feynman propagators, which take the form \cite{Bogolyubov:1980nc}
\begin{align}\label{eq:MinkPropagator}
	G_F(z) = \frac{m}{4\pi^2 \sqrt{-z^2}} K_1 (m\sqrt{-z^2}), 
\end{align}
where $K_1(\bullet)$ is the modified Bessel function of second kind, $z^2 = \et^\e_{\m\n} x^\m x^\n$ and $\et^\e_{\m\n} \bydef \diag(1-i\e,-1,-1,-1)$. We observe that Proposition \ref{pr:MinkIFF} is proved, using the result of Theorem \ref{th:ConstCoupLim}, if we find Euclidean bounds on the Feynman Propagator $G_F$. We obtain these bounds in two steps. First we rewrite Lemma \ref{le:MetricEstimate} for the Minkowski metric, which gives immediately the Euclidean bound on the first factor of \eqref{eq:MinkPropagator}. Second we establish Euclidean bounds directly on Bessel functions.
\begin{lemma}\label{le:arg}
	Let $\et^\e_{\m\n} \bydef \diag(1-i\e,-1,-1,-1)$, $\e>0$, the analytic continuation of the Minkowski metric. For 
	\begin{align}
		z^2 = \et^\e_{\m\n} x^\m x^\n = (1-i\e) x_0^2 - \bs{x}^2, \qquad x_E^2 = x_0^2 + \bs{x}^2
	\end{align}
	the following inequalities hold
	\begin{align}
		|\sqrt{-z^2}| & \leq (1+\e^2)^{\frac{1}{4}} |x_E| \bydef \check x_\e \label{eq:upperarg}\\
		|\sqrt{-z^2}| & \geq \left(\frac{1}{\e} + \sqrt{1+\frac{1}{\e^2}} \right)^{-\frac{1}{2}} |x_E| \bydef \hat x_\e .\label{eq:lowerarg}
	\end{align}
\end{lemma}
\begin{proof}
	The statement follows directly from Lemma \ref{le:MetricEstimate}.
\end{proof}
\begin{remark}
	It is important to note that we do not perform any kind of Wick rotation at any stage. 
\end{remark}
Next we want to derive bounds on the Bessel functions $K_1(\bullet)$. The functions $K_n(x)$ are positive and strictly monotonously decaying for $n\in\IN_0$ and $x\in\IR_+$. Hence we can control the decay properties of the propagator in every direction $e\in S^3\subset\IR^4$ if we manage to find estimates on $K_1(\bullet)$ with respect to the Euclidean norm in the sense of Lemma \ref{le:arg}.
\begin{proposition}\label{pr:bessel}
Let $\n\in \IN_0$, $\sqrt{-z^2} \in \IC$ and $|\arg (\sqrt{-z^2})| < \frac{\p}{2}$. Then 
\begin{align}
	K_\n(\check x_E) \leq K_\n(|\sqrt{-z^2}|) \leq |K_\n(\sqrt{-z^2})| \leq K_\n(\hat x_E)
\end{align}
with
\begin{align}
	\check x_E \stackrel{.}{=} & (1+\e^2)^{\frac{1}{4}} |x_E|\\
	\hat x_E \stackrel{.}{=} & \left(\frac{1}{\e} + \sqrt{1+\frac{1}{\e^2}} \right)^{-\frac{1}{2}} |x_E|.  
\end{align}
\end{proposition}
\begin{proof}
Consider the integral representation of modified Bessel functions of the second kind \cite{abramowitz1964handbook} for which we have to show that $|\arg (\sqrt{-z^2})| < \frac{\p}{2}$ or equivalently that $\Re(\sqrt{-z^2})>0$. This follows from Lemma \ref{le:arg} and it is straightforward to show the first inequality
\begin{align}
	K_\n(|\sqrt{-z^2}|) & = \int_{\IR^+}\limits dt\, e^{-|\sqrt{-z^2}| \cosh (t)} \cosh (\n t) \geq \int_{\IR^+}\limits dt\, e^{-\check x_E \cosh (t)} \cosh (\n t) = K_\n(\check x_E).
\end{align}
For the second estimate, recall that $K_\n(x) \in \IR^+$ for $x\in \IR^+$ and let $z\in\IC$. Then we have
\begin{align}
	K_\n(|z|) = |K_\n(|z|)|
\end{align}
and, using $z = |z| \cos(\vph) + i |z| \sin(\vph)$, estimate
\begin{align}
	K_\n(|z|) & = \int_{\IR^+}\limits e^{|z|\cosh(t)} \cosh(\n t) dt \\
	& \leq \int_{\IR^+}\limits e^{|z| \cos(\vph) \cosh(t)} \cosh(\n t) dt \\
	& = \Re \left\{ \int_{\IR^+}\limits e^{z \cosh(t)} \cosh(\n t) dt \right\}.
\end{align}
Since $|\Re(z')|\leq |\Re(z') + i \Im(z')| = |z'|$ for $z'\in \IC$, we obtain
\begin{align}
	\left| \Re \left\{ \int_{\IR^+}\limits e^{z \cosh(t)} \cosh(\n t) dt \right\} \right| \leq \left| \int_{\IR^+}\limits e^{z \cosh(t)} \cosh(\n t) dt \right|,
\end{align}
which proves the second estimate. Hence it remains to show the third estimate
\begin{align}
	|K_\n (\sqrt{-z^2})| & \leq \int_{\IR^+}\limits dt\, |e^{-\sqrt{-z^2} \cosh (t)} \cosh (\n t)| = \int_{\IR^+}\limits dt\, e^{-\Re(\sqrt{-z^2}) \cosh(t)} \cosh(\n t) \\
	& \leq \int_{\IR^+}\limits dt\, e^{-\hat x_E \cosh(t)} \cosh(\n t) = K_\n(\hat x_E), 
\end{align}
where we used Lemma \ref{le:arg} again.
\end{proof}    
Hence we obtain for the Feynman propagator
\begin{align}
G_F(\check x_E) \leq |G^\e_F(z)|\leq G_F(\hat x_E)
\end{align}
and thus for the kernel
\begin{align}
u_0^E[\G](\check x_{E,1},...,\check x_{E,|V(\G)|}) \leq |u^\e_0[\G](z_{1},...,z_{|V(\G)|})| \leq u_0^E[\G](\hat x_{E,1},...,\hat x_{E,|V(\G)|}).
\end{align}
This concludes the proof of Proposition \ref{pr:MinkIFF}. Since $G_F(x_E)$ is exponentially decaying for large arguments $x_E$ \cite{abramowitz1964handbook}, $\ird(G_F(x_E))$ is positive and goes to infinity. This fact implies an interesting special case of Theorem \ref{th:ConstCoupLim}.
\begin{corollary}\label{co:EachVertexOneMassiveLine}
	The limit of constant coupling exists if every internal vertex has one incident line, which corresponds to a propagator with positive mass parameter.
\end{corollary}
It is left to show that the $R$-operation does not decrease the decay behavior of the distribution kernel $u_0^\e[\G]$ so that assumption \eqref{eq:ConstCoupLimAssumption} is sufficient. Analogously to the proof of the convergence of the $R$-operation for short distance singularities, the following result establishes the desired relation.
\begin{theorem}[Thm. 2, \cite{Pottel:2017aa}]\label{th:L1}
	Let $\CK[\G]\in C^\infty(\IR^{d|V(\G)|}\setminus\oo)$ be the weight over a simple graph $\G$, which has positive scaling degree at the large graph diagonal. 
	Suppose that the IR-degree of divergence is positive for all $\CK[\g]$, $\g\subseteq\G$. Then
	\begin{align}
		R\CK[\G] \in L^1(\IR^{d|I|})
	\end{align} 
	for any $I\subset V(\G)$.
\end{theorem}
The proof of Theorem \ref{th:L1} builds upon Theorem \ref{th:L1loc}, i.e.\ it uses that $R\CK[\G] \in L^1_\mathrm{loc}(\IR^{d|I|})$.
Due to the local integrability, a Calderon-Zygmund-type decomposition can be constructed for $R\CK[\G]$, which allows to work with the forest formula \eqref{eq:forestformula} and does not require a reordering with respect to $I$.
With \cite{Lowenstein:1975rg}
\begin{align}
	\irs_I ( R \CK[\G] ) \geq \min_{F\in \SF} \irs_I \Big( \prod_{\g\in F} ( -t(\g) \IP(\g) \CK[\G] ) \Big) \, ,
\end{align}
the considerations can be further reduced to a generic $\G$-forest $F$ and it remains to show that the $R$-operation restricted to a single forest can only improve the scaling for large distances.
Analogously to the proof of Theorem \ref{th:L1loc}, which is described above, first the bound
\begin{align}
	\irs_{I_\la} ( -t(\g) \IP(\g) \CK[\G] ) \geq \irs_{I_\la} ( \CK[\la] )
\end{align}
is established for any $\la,\g \subset \G$ and then utilized to show that, for any $\g\in F$, $F\in \SF$, and $I\subset V(\G)$, the inequality
\begin{align}
	\irs_{I_\g} (R_{F,\g} \CK[\G]) \geq \irs_{I_\g} ( \CK[\g] )
\end{align}
holds, provided that the same relation holds for any renormalization part in $F$ which is also maximal subgraph of $\g$.
Starting the recursion from minimal renormalization parts of $F$ and proceeding to the maximal renormalization part, 
\begin{align}
	\irs_I ( R\CK[\G] ) \geq \irs_I ( \CK[\G] )
\end{align}
is obtained, which implies integrability of $R\CK[\G]$ over the vertex set $I$ since, by assumption, $\irs_I( \CK[\G] ) > d|I|$.

This concludes the proof of Theorem \ref{th:ConstCoupLim} as well. 
We observe that the limit of constant coupling can be performed for most time-ordered products. Obvious harmful settings are vertices, which have exactly two incident lines corresponding to propagators with vanishing mass parameter. From the estimates above, we obtain that those vertices induce an IR-scaling which is equal to the space dimension such that the IR-degree of divergence equals zero. We learn that time-ordered products, involving massless fields, require more attention in this limit. 

\section{Relation to the Momentum Space Scheme}

We established in Theorem \ref{th:Convergence} that the configuration space prescription defines a renormalization scheme and therefore must be equivalent to the original BPHZ scheme in momentum space.
It is evident though that both schemes cannot be related by simple Fourier transformation of the forest formula \eqref{eq:forestformula}.
This is most notable in the treatment of models with correlation functions obeying power-laws, e.g.\ if massless quantum fields are involved.
Then the Taylor subtractions at vanishing momenta may introduce new, unphysical IR-divergencies and require a modification of the prescription. 
In contrast, the IR-behavior of correlation functions may only be improved by the $R$-operation defined in configuration space according to Theorem \ref{th:ConstCoupLim}, which implies that massive and massless quantum fields can be treated on equal footing.
Further, we notice differences in the definition of renormalization parts, i.e.\ the definition via sets of edges versus sets of vertices or the constraint of one-particle irreducible graphs versus connected graphs.
In the following, we examine to which extend the components of the prescriptions can be related to each other. 

We find that, indeed, Zimmermann's choice of $\e$-regularization
\begin{align}
	p^2 - m^2 \mapsto p^2 - m^2 + i\e(\bs{p^2}+m^2) 
\end{align} 
corresponds to our choice of analytic continuation of the metric
\begin{align}
	\et \mapsto \et^\e = \diag (1-i\e,-1,-1,-1)
\end{align}
restricted to Minkowski space, and our choice of Taylor operation corresponds to Taylor polynomials at vanishing external momentum of the involved subgraph. 
The respective computations can be performed straightforwardly. 

It was pointed out in \cite{Brunetti:2009qc} (among others) that the additional subtractions in subgraphs required in the BPHZ momentum space scheme do not appear for the setting-sun diagram in Epstein-Glaser renormalization. This property was considered to be of advantage with respect to the BPHZ method in momentum space, and, indeed, it can be shown that these additional Taylor operations are redundant when proving equivalence to another renormalization scheme \cite{Zimmermann:1975gk}. However, the additional subtractions were required in momentum space for the absolute convergence of Feynman integrals, but may be omitted in the configuration space BPHZ version, too, due to the observation that the variables in momentum space are associated to lines and the variables in configuration space are associated to vertices. Note that the same momentum variable might appear as an argument in every line of the considered graph (with respect to admissible momentum flows given in \cite{Zimmermann:1969jj}). If one takes into account that the Taylor operation is performed in configuration space on all vertices of the graph, then the Taylor operator acting on a subgraph of the setting-sun graph has the same set of arguments as the Taylor operator acting on the full setting-sun graph. In general, this is not true for a chosen admissible flow in the momentum space approach regarding that the Taylor subtraction is performed on all external momenta, which may differ between the full setting-sun graph and any subgraph. 

Associated to each formulation of BPHZ renormalization, we want to treat the additional subtractions, which have to be performed due to the appearance of additional renormalization parts, separately. Already in the proof of Theorem \ref{th:Convergence} we performed the transition from Feynman graphs to simple graphs. This corresponds to neglecting divergent subgraphs of renormalization parts, where the former have the same set of vertices but less edges, which implies a deviation in the point of Taylor subtraction using \eqref{eq:CMCoordinate}. In the momentum space prescription these divergent subgraphs have to be taken into account because they are assigned to a subset of free loop integrations. However, taking those into account in the configuration space approach differs from the result in Theorem \ref{th:Convergence} only by combinatorial factors.
\begin{proposition}\label{pr:SubgraphCombinatorics}
	Given a multigraph $\G$ and subgraphs $\g'\subset\g\subseteq \G$ with $V(\g)=V(\g')$ and $d(\g)>d(\g')\geq 0$.
	Then
	\begin{align}
		t(\g) t(\g') u_0^\e[\G] = u_0^\e[\g] \sum_{|\al|=0}^{d(\g)} C(\al) \frac{(x-\ol x_\g)^\al}{\al!}  D^\al_{V(\g)|\ol{V(\g)}} u_0^\e[\G \lineco \g] 
	\end{align}
	and
	\begin{multline}
		(1-t(\g)) (1-t(\g')) u_0^\e[\G] \\
		= u_0^\e[\g] C' \sum_{|\al|=d(\g)+1} C''(\al) \frac{(x-\ol x_\g)^\al}{\al!} D^\al_{V(\g)|\ol{V(\g)}} u_0^\e[\G \lineco \g] \, ,
	\end{multline}
	where $C$, $C'$ and $C''$ are combinatorial factors depending on $d(\g)$, $d(\g')$, and the dimension of the spacetime $d$ as well as the form of the Schl\"omilch remainder.
\end{proposition}
\begin{proof}
	Let $\g'\subset\g\subseteq \G$ with $V(\g)=V(\g')$ and $d(\g)>d(\g')\geq 0$. 
	Then we have to distinguish two cases. Either we have to compute the Taylor remainder or the Taylor polynomial for both subgraphs. We begin with the latter. Note that since $\g'\subset \g$, $t(\g')$ is applied before $t(\g)$ but due to $V(\g)=V(\g')$ they are applied to the same set of variables, i.e.\
	\begin{align}
		t^{d(\g)}_{V(\g)|\ol{V(\g)}} t^{d(\g')}_{V(\g')|\ol{V(\g')}} & = \sum_{|\al|=0}^{d(\g)} \frac{(x-\ol x)^\al}{\al!} D^\al_{x|\ol x} \sum_{|\be|=0}^{d(\g')} \frac{(x-\ol x')^\be}{\be!} D^\be_{x|\ol x'} \\
		& = \sum_{|\al|=0}^{d(\g)} \sum_{|\be|=0}^{d(\g')} \frac{(x-\ol x)^\al}{\al!} D^\al_{x|\ol x} \frac{(x-\ol x')^\be}{\be!} D^\be_{x|\ol x'} \\
		& = \sum_{|\al|=0}^{d(\g)} \sum_{\be\subseteq\al} \frac{(x-\ol x)^\al}{\al!} D^\al_{x|\ol x} \\
		& = \sum_{|\al|=0}^{d(\g)} C(\al) \frac{(x-\ol x)^\al}{\al!}  D^\al_{x|\ol x}.
	\end{align}
	We used that the expression vanishes if $\be\nsubseteq\al$. For the remainder term, we consider a function $f\in C^k(\IR^n\times \IR^m)$ with $k>d(\g')+d(\g)+1$ and write the first application of $(1-t(\g'))$ in Schl\"omilch form
	\begin{align}
		(1-t^{d(\g')}_{V(\g')|\ol{V(\g')}})f(x,y) & = (1-t^{d(\g')}_{x|\ol x'})f(x,y) \\
		& = \frac{(d(\g')+1)(1-\vte)^{d(\g')+1-p'}}{p'}\times \\
		& \hspace{0.5cm} \times \sum_{|\be|=d(\g')+1} \frac{(x-\ol x')^\be}{\be!} D^\be_x f(\ol x' +\vte (x-\ol x'),y) \\
		& = C'(d(\g'),p',\vte) \times \\
		& \hspace{0.5cm} \times \sum_{|\be|=d(\g')+1} \frac{(x-\ol x')^\be}{\be!} D^\be_x f(\ol x' +\vte (x-\ol x'),y) \\ & = R_{d(\g')}f(x,y;\ol x').
	\end{align}
	Instead of calculating the second remainder ``on top'' of the first one, we formally expand
	\begin{align}\label{eq:ExpandRf}
		R_{d(\g')}f(x,y;\ol x') = \sum_{|\g|=0}^k \frac{(x- \ol x')^\g}{\g!} D^\g_{x|\ol x'} R_{d(\g')}f(x,y;\ol x') 
	\end{align}
	and compute the Taylor polynomial
	\begin{align}\label{eq:TayRf}
		t^{d(\g)}_{V(\g)|\ol{V(\g)}} R_{d(\g')}f(x,y;\ol x') = \sum_{|\al|=0}^{d(\g)} \frac{(x- \ol x)^\al}{\al!} D^\al_{x|\ol x} R_{d(\g')}f(x,y;\ol x').
	\end{align}
	We observe that for $|\al|\geq d(\g')+1$ 
	\begin{multline}\label{eq:DonRf}
		D^\al_{x|\ol x} \left( C'(d(\g'),p',\vte) \sum_{|\be|=d(\g')+1} \frac{(x-\ol x')^\be}{\be!} D^\be_x f(\ol x' +\vte (x-\ol x'),y) \right) \\
		= C'(d(\g'),p',\vte) C''(\al,\vte) f^{(\al)}(\ol x,y),
	\end{multline}
	where $C(\al,\vte)$ is the same combinatorial factor as above except for the additional factors $\vte$ coming from the chain rule. Subtracting \eqref{eq:TayRf} from \eqref{eq:ExpandRf} and using \eqref{eq:DonRf}, we arrive at
	\begin{multline}
		(1-t^{d(\g)}_{V(\g)|\ol{V(\g)}}) R_{d(\g')}f(x,y;\ol x') \\ = \sum_{|\al|=d(\g)+1} \frac{(x-\ol x)^\al}{\al!} C'(d(\g'),p',\vte) C''(\al,\vte) f^{(\al)}(\ol x,y),
	\end{multline}
	which, apart from the combinatorial factors $C'$ and $C''$, has the desired property with respect to the $R$-operation.
\end{proof}
In the forest formula, renormalization parts are defined to be connected full vertex parts of (sub-)graphs. These introduce again additional subtractions to other schemes, e.g. Epstein-Glaser scheme or analytic regularization and, in particular, the BPHZ scheme in momentum space. The reason for this deviation lies in the fact that, for configuration space treatments like Epstein-Glaser or the analytic regularization, graphs remain affected by the regularization of subgraphs. This sustained effect is a result of choosing a regularization, which does not modify the graph weight itself. It does not hold for our formulation of BPHZ renormalization in configuration space, where one cannot expect an improvement of the scaling behavior of graphs induced by Taylor operations on subgraphs. Instead, for the forest formula of the momentum space method, only proper (or one-particle-irreducible) graphs were considered. The reasoning is that divergent contributions in momentum space stem from integrations over free internal momenta in closed loops. Therefore the Taylor operations acting on non-proper (or one-particle-reducible) graphs would translate to the Taylor expansion around vanishing momenta of the Fourier transform. But any polynomial is a well-defined Schwartz-distribution, thus amounts to a finite change of the considered quantity, e.g. an $S$-matrix element, provided the weight is defined for exceptional momenta. We conclude that momentum space BPHZ renormalization is related to our configuration space BPHZ prescription by Fourier transformation if all renormalization parts of both schemes are considered in the forest formula. Combinatorial factors and finite changes coming from additional subtractions in one or the other scheme are compensated after employing appropriate normalization conditions. Note that this relation can only hold for fields with positive mass parameter, since the momentum space method is generally defined only for those. 

For fields with vanishing mass parameter, the results of Theorem \ref{th:Convergence} hold in configuration space, but are doomed to fail after performing the Fourier transformation of the forest formula. Nevertheless we may examine the properties of the inverse Fourier transformation for the modified BPHZ scheme in momentum space. In the original works \cite{Lowenstein:1974qt,Lowenstein:1975rf,Lowenstein:1975rg,Lowenstein:1975ps,Lowenstein:1975ku}, Lowenstein and Zimmermann introduced an auxiliary mass term $M(1-s)$ with $s\in[0,1]$ in order to deal with spurious infrared divergences that appear in the Taylor subtractions of the BPHZ scheme for massless theories. The strategy is basically to ``double'' the Taylor operation by another one with an IR-subtraction degree $r(\g)$ or in case of oversubtraction $\rh(\g)\leq r(\g)$. These degrees are a bit misleading, as they do neither make a reference to the actual IR-scaling of the distribution nor do they describe an actual change in the IR-scaling. In momentum space, it was defined as
\begin{align}\label{eq:pTaylorOp}
	1-\hat\ta^{r,d}_{p,s}(\g) \bydef (1-\hat t^{r(\g)-1}_{p,s-1}) (1-\hat t^{d(\g)}_{p,s}),
\end{align}
where 
\begin{align}
	t^d_{x,y,z...} = t^d_{x=0,y=0,z=0,...} \qquad \mbox{for } d\geq 0
\end{align}
and $0$ otherwise. There are additionally some relations among the degrees and rules for oversubtractions, which we do not discuss here. It is important to recall the strategy of Lowenstein and Zimmermann. The first subtractions with degree $d(\g)$ are performed at positive mass $M$, such that no additional IR-divergences get introduced into the considered integrand. Then the second Taylor subtraction is performed at vanishing auxiliary mass in order to restore the correct normalization for massless propagators (and 3-point functions). We emphasize that the second subtractions are not performed to improve the IR-behavior of the considered integrand.

It is necessary to discuss the role of the parameter $s$ in Fourier transformation. Lowenstein and Zimmermann treated it as a variable taking part in the scaling. While this seems reasonable for rational functions, where the $s$-dependent terms appeared additive to the momentum space variables, i.e.\
\begin{align}
	p^2- M^2(s-1)^2 + i\e(\bs{p}^2+M^2(s-1)^2), 
\end{align} 
we are facing a multiplicative dependence in position space, i.e.\ we find the argument
\begin{align}
	M(1-s)\sqrt{-x^2_\e}.
\end{align} 
Further we do not consider $s$ to be a variable keeping in mind that we change the differential operator $P$ by a constant potential and not by an additional variable that would extend the configuration space to another dimension. Instead, we want to understand the interplay of $x$- and $s$-derivatives in the Taylor operators. For this purpose, we start with a generic
\begin{align}
	t^m_{x,s|\ol x,\ol s} \bydef \sum^m_{|\al|+a=0} \frac{(x-\ol x)^\al (s-\ol s)^a}{\al!a!} D^\al_{x|\ol x} D^a_{s|\ol s},
\end{align}
where $\al$ is a multi-index and $a$ is an integer. Turning towards the Fourier transform of \eqref{eq:pTaylorOp}, i.e.\
\begin{align}
	1-\ta^{r,d}_{x,s}(\g) \bydef (1- t^{d(\g)}_{x,s|\ol x,0}) (1- t^{r(\g)-1}_{x,s|\ol x, 1}),
\end{align}
we observe that its definition involves the product of Taylor operators
\begin{align}
	t^{d(\g)}_{x,s|\ol x,0}t^{r(\g)-1}_{x,s|\ol x, 1} = \sum^{d(\g)}_{|\al|+a=0} \frac{(x-\ol x)^\al s^a}{\al!a!} D^\al_{x|\ol x} D^a_{s|0} \sum^{r(\g)-1}_{|\be|+b=0} \frac{(x-\ol x)^\be (s-1)^b}{\be!b!} D^\be_{x|\ol x} D^b_{s|1},
\end{align}
which leads immediately to the conditions $a\leq b$ and $\al\supseteq\be$ for non-vanishing contributions. A crucial difference to the momentum space treatment of Lowenstein and Zimmermann is the relation of the variables $s$ and $x$ in the propagator. In order to illustrate this, consider a sufficiently smooth function $f\in C^k(\IR^d\times [0,1])$ and some Taylor operator $t^m_{x,s|\ol x, \ol s}$. First we compute the $s$-derivative for the case $f(x,s)=f(s\cdot g(x))$, where $g(x)$ is some nonlinear function of dimension 1 in $x$, and check the scaling. We get $\pa_{s|\ol x,\ol s} f(x,s) = g(\ol x) f'(\ol s,\ol x)$ for the Taylor operator and do not observe any change in the UV-scaling assuming that $f$ is smooth in a neighborhood of $\ol x$. While this does not pose any issue for ``bare'' Taylor operators acting on the distribution kernel as we saw in the proof of Theorem \ref{th:L1loc}, treating Taylor remainder is problematic for the same reason. For notational simplicity, we perform the calculation in several steps. In Schl\"omilch form, we write
\begin{align}
	(1-t^m_{x,s|\ol x, \ol s}) f(x,y;s) & = \frac{(m+1)(1-\te)^{m+1-p}}{p} \sum_{|\al|+a=m+1} \frac{(x-\ol x)^\al (s-\ol s)^a}{\al! a!} \times \nonumber\\
	& \qquad\qquad\qquad\qquad \times D^{\al,a} f(\ol x + \te (x-\ol x),y, \ol s + \te(s-\ol s)) \\
	& \bydef R_mf(x,y,\ol x, \te;s, \ol s)
\end{align}
Taking the Taylor remainder again, but with at least equal order $n\geq m$ and at the point $(\ol x, \ol s')$, we arrive at
\begin{multline}
	(1-t^n_{x,s|\ol x, \ol s'}) R_mf(x,y,\ol x, \te;s, \ol s) = \\
	= \frac{(n+1)(1-\te')^{n+1-p'}}{p'} \sum_{|\be|+b=n+1} \frac{(x-\ol x)^\be (s-\ol s')^b}{\be! b!} \times \\ \times D^{\be,b} R_mf(\ol x + \te' (x-\ol x),y,\ol x, \te;\ol s' +\te'(s-\ol s'), \ol s).
\end{multline}
In the extremal case of having $a=m+1$ and $b=n+1$, we are left without moments of the form $(x-\ol x)$ in the remainder, which were crucial for the lowering of the UV-scaling degree. Adding the analysis on $s$-derivatives, we find moments of the form
\begin{align}
	g(\ol x + \te(x-\ol x) - \ol x,y) = \sqrt{(\te(x-\ol x) - y)^2} ,
\end{align}
which attain the value $y$ in the scaling $x\rightarrow \ol x$. Hence we cannot expect an improvement in the sense of a smaller UV-scaling degree from $s$-derivatives and we obtain formally
\begin{align}
	\uvs & \left(  R_nR_mf(x,\ol x,y,\te;s,\ol s,\ol s') \right) \\
	& \leq \uvs \Bigg( \sum_{|\be|+b=n+1} \frac{(x-\ol x)^\be (s-\ol s')^b}{\be! b!} \times \\
	& \hspace{2.5cm} \times D^{\be,b} R_mf(\ol x + \te' (x-\ol x),y,\ol x, \te;\ol s' +\te'(s-\ol s'), \ol s) \Bigg) \\
	& \leq \max_{|\be|+b=n+1} \uvs\Bigg( \frac{(x-\ol x)^\be (s-\ol s')^b}{\be! b!} \times \\
	& \hspace{2.5cm} \times D^{\be,b} R_mf(\ol x + \te' (x-\ol x),y,\ol x, \te;\ol s' +\te'(s-\ol s'), \ol s) \Bigg) \\
	& \leq \uvs\left( D^{0,n+1} R_mf(\ol x + \te' (x-\ol x),y,\ol x, \te;\ol s' +\te'(s-\ol s'), \ol s) \right) \\
	& \leq \uvs\left( R_mf(\ol x + \te' (x-\ol x),y,\ol x, \te;\ol s' +\te'(s-\ol s'), \ol s) \right) \\
	& \leq \uvs \left( f(\ol x + \te' (x-\ol x),y,\ol x, \te;\ol s' +\te'(s-\ol s'), \ol s) \right),
\end{align}
where one repeats the arguments in the last step. Since we assumed that $f$ is sufficiently smooth, we have $\uvs(f)< d$ but no further improvement which is required for the UV-convergence. Therefore this result indicates that, computing the naive inverse Fourier transformation of the forest formula, the BPHZL method does not define a renormalization scheme in configuration space. In the same sense, our configuration space method does not define renormalization scheme for massless scalar fields in momentum space after naive Fourier transformation of the forest formula, since we established the relation already to the BPHZ prescription. However, this is no contradiction to the equivalence of renormalization schemes, because we related only the regularization prescriptions, i.e.\ the Taylor operators in the forest formula, and did not attempt to prove equivalence. 
Nevertheless, the fact that the configuration space approach is agnostic towards mass parameters of quantum fields may be interpreted as an advantage of the scheme in comparison to its counterpart in momentum space.

\section{Conclusions}

In the present work, the concept of BPHZ renormalization is extended to analytic spacetimes using the algebraic approach to perturbative quantum field theory. We shall describe how the construction in configuration space arises naturally maintaining contact to the notions of the original works.\\ 
First, the larger class of spacetime geometries does generally not admit a global treatment in momentum space so that all arguments are formulated entirely in configuration space. The change in fundamental variables, from momentum flow through edges of Feynman graphs to loci of vertices in spacetime, amends the problem under consideration to local integrability or existence of generalized convolutions in small regions of spacetime. Therefore Theorem \ref{th:Convergence} is formulated only for geodesically convex regions, which is indeed sufficient to solve the extension problem connected to renormalization. However, it may be extended to the whole spacetime by a partition of unity argument \cite{Moretti:2001qh} or reformulated using quasifree states, which admit a definition of correlation functions in terms of Feynman graphs on the whole spacetime. In general, the issue of scheme-compatible states, for instance thermal equilibrium states \cite{Fredenhagen:2013cna}, is left for future research and might require a generalized notion of Feynman graphs, for instance $\De$-complexes of higher dimension \cite{hatcher2002algebraic}, in the latter case. \\
Second, our construction is agnostic towards a mass parameters of quantum fields. 
Comparing this to the findings of BPHZ and BPHZL renormalization in momentum space, it is impossible the all three methods are related by Fourier transformation on the level of the forest formula. 
Recall that the modification of the BPHZ method was required if massless fields were present, so that we could expect to establish a relation only to one if any momentum space approach. Indeed, we show that our prescription can be transferred (up to combinatorial factors) into the BPHZ momentum space method for positive masses. 
In contrast, the BPHZL approach does not define a renormalization scheme in configuration space after inverse Fourier transformation of the forest formula. 
We remark that this does not pose a contradiction regarding the equivalence of renormalization schemes, but expresses just that prescriptions are tailored according to the conditions at hand.
In fact, we shall view the treatment of massive and massless quantum fields on equal footing as a benefit of the configuration space approach.\\
Finally, choosing the subtraction point with edge set weighted coordinates over the subtraction point with uniformly weighted coordinates shows its relevance in the derivation of normal products and Zimmermann identities \cite{Pottel:2017cc}. Namely, the properties first come into play, where non-trivial manipulations of the graphs are performed so that the renormalization scheme can be proved and the relation to the momentum space method can be established using either definition. However, the coincidence limit of vertices in graphs can only be described by the edge set weighted version.

It is worth noting that, like in the original work of Zimmermann, the scheme is formulated and proved on the level of weighted Feynman graphs, which may be summed up to time-ordered products of, in our approach, a finite number of Wick monomials. Relaxing this to include time-ordered products of possibly infinitely many monomials, one may use our results to study structural properties of concrete theories.
This would most likely demand a reformulation of the action principle \cite{Lowenstein:1971jk} so that parametric differential equations \cite{Zimmermann:1979fd,Hollands:2002ux,Brunetti:2009qc} or the behavior of physical quantities under symmetry transformations \cite{Kraus:1991cq,Kraus:1992ru} should be derivable more conveniently. Furthermore it would be interesting to investigate the interplay of our renormalization prescription with the BRST- \cite{Hollands:2007zg} or BV-formalism \cite{Fredenhagen:2011mq} in regard to vector fields as well as studying supersymmetric extensions \cite{Piguet:1986ug} of quantum field theories or theories over non-commutative spacetime \cite{Blaschke:2013cba}.
\subsection*{Acknowledgments}

The author would like to thank Klaus Sibold for numerous discussions. 
Parts of this work have been concluded during a stay at Leipzig University and the Max Planck Institute for Mathematics in the Sciences.
Their hospitality and the financial support by the International Max Planck Research School (IMPRS) ``Mathematics in the Sciences'' is gratefully acknowledged. 

\bibliographystyle{../halpha}
\bibliography{../bphzl}

\newcommand{\etalchar}[1]{$^{#1}$}
\begin{thebibliography}{BGH{\etalchar{+}}13}

\bibitem[AS64]{abramowitz1964handbook}
M.~Abramowitz and I.A. Stegun.
\newblock {\em Handbook of Mathematical Functions: With Formulas, Graphs, and
  Mathematical Tables}.
\newblock Applied mathematics series. Dover Publications, 1964.

\bibitem[BDF09]{Brunetti:2009qc}
R.~Brunetti, M.~Duetsch, and K.~Fredenhagen.
\newblock {Perturbative Algebraic Quantum Field Theory and the Renormalization
  Groups}.
\newblock {\em Adv.Theor.Math.Phys.}, 13:1541--1599, 2009, 0901.2038.

\bibitem[BF00]{Brunetti:1999jn}
R.~Brunetti and K.~Fredenhagen.
\newblock {Microlocal analysis and interacting quantum field theories:
  Renormalization on physical backgrounds}.
\newblock {\em Commun.Math.Phys.}, 208:623--661, 2000, math-ph/9903028.

\bibitem[BFV03]{Brunetti:2001dx}
R.~Brunetti, K.~Fredenhagen, and R.~Verch.
\newblock {The Generally covariant locality principle: A New paradigm for local
  quantum field theory}.
\newblock {\em Commun. Math. Phys.}, 237:31--68, 2003, math-ph/0112041.

\bibitem[BG72]{Bollini:1972ui}
C.~G. Bollini and J.~J. Giambiagi.
\newblock {Dimensional Renormalization: The Number of Dimensions as a
  Regularizing Parameter}.
\newblock {\em Nuovo Cim.}, B12:20--26, 1972.

\bibitem[BGH{\etalchar{+}}13]{Blaschke:2013cba}
D.~N. Blaschke, F.~Gieres, F.z Heindl, M.~Schweda, and M.~Wohlgenannt.
\newblock {BPHZ renormalization and its application to non-commutative field
  theory}.
\newblock {\em Eur. Phys. J.}, C73:2566, 2013, 1307.4650.

\bibitem[BGP07]{Bar:2007zz}
C.~Bar, N.~Ginoux, and F.~Pfaffle.
\newblock {\em {Wave equations on Lorenzian manifolds and quantization}}.
\newblock 2007.

\bibitem[BP57]{Bogoliubov:1957gp}
N.~N. Bogoliubov and O.~S. Parasiuk.
\newblock {On the Multiplication of the causal function in the quantum theory
  of fields}.
\newblock {\em Acta Math.}, 97:227--266, 1957.

\bibitem[BRS76]{Becchi:1975nq}
C.~Becchi, A.~Rouet, and R.~Stora.
\newblock {Renormalization of Gauge Theories}.
\newblock {\em Annals Phys.}, 98:287--321, 1976.

\bibitem[BS59]{Bogolyubov:1980nc}
N.N. Bogolyubov and D.V. Shirkov.
\newblock {Introduction to the Theory of Quantized Fields}.
\newblock {\em Intersci.Monogr.Phys.Astron.}, 3:1--720, 1959.

\bibitem[CL76]{Clark:1976ym}
T.~E. Clark and J.~H. Lowenstein.
\newblock {Generalization of Zimmermann's Normal-Product Identity}.
\newblock {\em Nucl. Phys.}, B113:109--134, 1976.

\bibitem[DFKR14]{Duetsch:2013xca}
M.~Duetsch, K.~Fredenhagen, K.~J. Keller, and K.~Rejzner.
\newblock {Dimensional Regularization in Position Space, and a Forest Formula
  for Epstein-Glaser Renormalization}.
\newblock {\em J. Math. Phys.}, 55:122303, 2014, 1311.5424.

\bibitem[Dys49]{Dyson:1949bp}
F.~J. Dyson.
\newblock {The Radiation theories of Tomonaga, Schwinger, and Feynman}.
\newblock {\em Phys. Rev.}, 75:486--502, 1949.

\bibitem[EG73]{Epstein:1973gw}
H.~Epstein and V.~Glaser.
\newblock {The Role of locality in perturbation theory}.
\newblock {\em Annales Poincare Phys.Theor.}, A19:211--295, 1973.

\bibitem[FL14]{Fredenhagen:2013cna}
K.~Fredenhagen and F.~Lindner.
\newblock {Construction of KMS States in Perturbative QFT and Renormalized
  Hamiltonian Dynamics}.
\newblock {\em Commun. Math. Phys.}, 332(3):895--932, 2014, 1306.6519.

\bibitem[FR13]{Fredenhagen:2011mq}
K.~Fredenhagen and K.~Rejzner.
\newblock {Batalin-Vilkovisky formalism in perturbative algebraic quantum field
  theory}.
\newblock {\em Commun. Math. Phys.}, 317:697--725, 2013, 1110.5232.

\bibitem[Fri10]{Friedlander:2010eqa}
F.~G. Friedlander.
\newblock {\em {The Wave Equation on a Curved Space-Time}}.
\newblock Cambridge University Press, 2010.

\bibitem[GHP16]{Gere:2015qsa}
A.~G\'{e}r\'{e}, T.-P. Hack, and N.~Pinamonti.
\newblock {An analytic regularisation scheme on curved space–times with
  applications to cosmological space–times}.
\newblock {\em Class. Quant. Grav.}, 33(9):095009, 2016, 1505.00286.

\bibitem[Hat02]{hatcher2002algebraic}
A.~Hatcher.
\newblock {\em Algebraic Topology}.
\newblock Cambridge University Press, 2002.

\bibitem[Hep66]{Hepp:1966eg}
K.~Hepp.
\newblock {Proof of the Bogolyubov-Parasiuk theorem on renormalization}.
\newblock {\em Commun.Math.Phys.}, 2:301--326, 1966.

\bibitem[Hep69]{Hepp:1969bn}
K.~Hepp.
\newblock {On the equivalence of additive and analytic renormalization}.
\newblock {\em Commun. Math. Phys.}, 14:67--69, 1969.

\bibitem[Hol08]{Hollands:2007zg}
S.~Hollands.
\newblock {Renormalized Quantum Yang-Mills Fields in Curved Spacetime}.
\newblock {\em Rev. Math. Phys.}, 20:1033--1172, 2008, 0705.3340.

\bibitem[Hol13]{Hollands:2010pr}
S.~Hollands.
\newblock {Correlators, Feynman diagrams, and quantum no-hair in deSitter
  spacetime}.
\newblock {\em Commun. Math. Phys.}, 319:1--68, 2013, 1010.5367.

\bibitem[H{\"o}r90]{hormander1990analysis}
L.~H{\"o}rmander.
\newblock {\em The analysis of linear partial differential operators:
  Distribution theory and Fourier analysis}.
\newblock Springer Study Edition. Springer-Verlag, 1990.

\bibitem[HW01]{Hollands:2001nf}
S.~Hollands and R.~M. Wald.
\newblock {Local Wick polynomials and time ordered products of quantum fields
  in curved space-time}.
\newblock {\em Commun.Math.Phys.}, 223:289--326, 2001, gr-qc/0103074.

\bibitem[HW02]{Hollands:2001fb}
S.~Hollands and R.~M. Wald.
\newblock {Existence of local covariant time ordered products of quantum fields
  in curved space-time}.
\newblock {\em Commun.Math.Phys.}, 231:309--345, 2002, gr-qc/0111108.

\bibitem[HW03]{Hollands:2002ux}
S.~Hollands and R.~M. Wald.
\newblock {On the renormalization group in curved space-time}.
\newblock {\em Commun. Math. Phys.}, 237:123--160, 2003, gr-qc/0209029.

\bibitem[HW05]{Hollands:2004yh}
S.~Hollands and R.~M. Wald.
\newblock {Conservation of the stress tensor in interacting quantum field
  theory in curved spacetimes}.
\newblock {\em Rev. Math. Phys.}, 17:227--312, 2005, gr-qc/0404074.

\bibitem[IZ80]{Itzykson:1980rh}
C.~Itzykson and J.~B. Zuber.
\newblock {\em {Quantum Field Theory}}.
\newblock International Series In Pure and Applied Physics. McGraw-Hill, New
  York, 1980.

\bibitem[KS92]{Kraus:1991cq}
E.~Kraus and K.~Sibold.
\newblock {Conformal transformation properties of the energy momentum tensor in
  four-dimensions}.
\newblock {\em Nucl. Phys.}, B372:113--144, 1992.

\bibitem[KS93]{Kraus:1992ru}
E.~Kraus and K.~Sibold.
\newblock {Local couplings, double insertions and the Weyl consistency
  condition}.
\newblock {\em Nucl. Phys.}, B398:125--154, 1993.

\bibitem[KW91]{Kay:1988mu}
B.~S. Kay and R.~M. Wald.
\newblock {Theorems on the Uniqueness and Thermal Properties of Stationary,
  Nonsingular, Quasifree States on Space-Times with a Bifurcate Killing
  Horizon}.
\newblock {\em Phys. Rept.}, 207:49--136, 1991.

\bibitem[Low71]{Lowenstein:1971jk}
J.~H. Lowenstein.
\newblock {Differential vertex operations in Lagrangian field theory}.
\newblock {\em Commun. Math. Phys.}, 24:1--21, 1971.

\bibitem[Low76]{Lowenstein:1975ps}
J.~H. Lowenstein.
\newblock {Convergence Theorems for Renormalized Feynman Integrals with
  Zero-Mass Propagators}.
\newblock {\em Commun. Math. Phys.}, 47:53--68, 1976.

\bibitem[LS76]{Lowenstein:1975ku}
J.~H. Lowenstein and E.~R. Speer.
\newblock {Distributional Limits of Renormalized Feynman Integrals with
  Zero-Mass Denominators}.
\newblock {\em Commun. Math. Phys.}, 47:43--51, 1976.

\bibitem[LZ75a]{Lowenstein:1974qt}
J.~H. Lowenstein and W.~Zimmermann.
\newblock {On the Formulation of Theories with Zero Mass Propagators}.
\newblock {\em Nucl. Phys.}, B86:77--103, 1975.

\bibitem[LZ75b]{Lowenstein:1975rg}
J.~H. Lowenstein and W.~Zimmermann.
\newblock {The Power Counting Theorem for Feynman Integrals with Massless
  Propagators}.
\newblock {\em Commun. Math. Phys.}, 44:73--86, 1975.

\bibitem[LZ76]{Lowenstein:1975rf}
J.~H. Lowenstein and W.~Zimmermann.
\newblock {Infrared Convergence of Feynman Integrals for the Massless $A^4$
  Model}.
\newblock {\em Commun. Math. Phys.}, 46:105--118, 1976.

\bibitem[Mor03]{Moretti:2001qh}
V.~Moretti.
\newblock {Comments on the stress energy tensor operator in curved space-time}.
\newblock {\em Commun. Math. Phys.}, 232:189--221, 2003, gr-qc/0109048.

\bibitem[Pot17a]{Pottel:2017aa}
S.~Pottel.
\newblock {A BPHZ Theorem in Configuration Space}, 2017, 1706.06762.

\bibitem[Pot17b]{Pottel:2017cc}
S.~Pottel.
\newblock {Normal Products and Zimmermann Identities in Configuration Space
  BPHZ Renormalization}, 2017, 1708.04115.

\bibitem[PS86]{Piguet:1986ug}
O.~Piguet and K.~Sibold.
\newblock {\em {Renormalized Supersymmetry. The Perturbation Theory of N=1
  Supersymmetric Theories in Flat Space-Time}}.
\newblock Progress In Physics, 12. Boston, Usa: Birkhaeuser, 1986.

\bibitem[Spe71]{Speer:1972wz}
E.~R. Speer.
\newblock {On the structure of analytic renormalization}.
\newblock {\em Commun. Math. Phys.}, 23:23--36, 1971.
\newblock [Erratum: Commun. Math. Phys.25,336(1972)].

\bibitem[Ste00]{Steinmann:2000nr}
O.~Steinmann.
\newblock {\em {Perturbative quantum electrodynamics and axiomatic field
  theory}}.
\newblock 2000.

\bibitem[tHV72]{'tHooft:1972fi}
G.~'t~Hooft and M.~J.~G. Veltman.
\newblock {Regularization and Renormalization of Gauge Fields}.
\newblock {\em Nucl. Phys.}, B44:189--213, 1972.

\bibitem[Tyu75]{Tyutin:1975qk}
I.~V. Tyutin.
\newblock {Gauge Invariance in Field Theory and Statistical Physics in Operator
  Formalism}.
\newblock 1975, 0812.0580.

\bibitem[Vel76]{Velo:1976gh}
Velo, G. and Wightman, A. S.
\newblock {\em {Renormalization Theory. Proceedings, NATO Advanced Study
  Institute}}, volume~23 of {\em NATO ASI, C - Mathematical and Physical
  Sciences}, Dordrecht, 1976. Springer.

\bibitem[WZ72]{Wilson:1972ee}
K.~G. Wilson and W.~Zimmermann.
\newblock {Operator product expansions and composite field operators in the
  general framework of quantum field theory}.
\newblock {\em Commun. Math. Phys.}, 24:87--106, 1972.

\bibitem[Zim68]{Zimmermann:1968mu}
W.~Zimmermann.
\newblock {The power counting theorem for minkowski metric}.
\newblock {\em Commun.Math.Phys.}, 11:1--8, 1968.

\bibitem[Zim69]{Zimmermann:1969jj}
W.~Zimmermann.
\newblock {Convergence of Bogolyubov's method of renormalization in momentum
  space}.
\newblock {\em Commun. Math. Phys.}, 15:208--234, 1969.

\bibitem[Zim73a]{Zimmermann:1972te}
W.~Zimmermann.
\newblock {Composite operators in the perturbation theory of renormalizable
  interactions}.
\newblock {\em Annals Phys.}, 77:536--569, 1973.

\bibitem[Zim73b]{Zimmermann:1972tv}
W.~Zimmermann.
\newblock {Normal products and the short distance expansion in the perturbation
  theory of renormalizable interactions}.
\newblock {\em Annals Phys.}, 77:570--601, 1973.

\bibitem[Zim75]{Zimmermann:1975gk}
W.~Zimmermann.
\newblock {Remark on Equivalent Formulations for Bogolyubov's Method of
  Renormalization}.
\newblock In {\em {Renormalization Theory. Proceedings, NATO Advanced Study
  Institute: Erice, 17-31 August, 1975}}, pages 161--170, 1975.

\bibitem[Zim80]{Zimmermann:1979fd}
Wolfhart Zimmermann.
\newblock {The Renormalization Group of the Model of $A^4$ Coupling in the
  Abstract Approach of Quantum Field Theory}.
\newblock {\em Commun. Math. Phys.}, 76:39, 1980.

\end{thebibliography}

\end{document}